\documentclass[11pt,twoside]{article}
\usepackage{amsfonts,amssymb,stmaryrd,amsthm}
\usepackage{amsbsy}
\usepackage{graphicx}
\usepackage{latexsym}
\usepackage{psfrag}
\usepackage{subfigure}
\usepackage{amsmath,mathrsfs,bm}

\begin{document}
\sloppy
\newcommand{\dickebox}{{\vrule height5pt width5pt depth0pt}}

\newtheorem{Def}{Definition}[section]
\newtheorem{Eg}[Def]{Example}
\newtheorem{Prop}[Def]{Proposition}
\newtheorem{Thm}[Def]{Theorem}
\newtheorem{Lem}[Def]{Lemma}
\newtheorem{Rem}[Def]{Remark}
\newtheorem{Coro}[Def]{Corollary}
\newtheorem{Conv}[Def]{Convention}

\newcommand{\ol}{\overline}
\newcommand{\ah}{\alpha}
\newcommand{\bt}{\beta}
\newcommand{\lra}{\longrightarrow}
\newcommand{\os}{\overset}
\newcommand{\us}{\underset}
\newcommand{\sm}{\sigma}
\newcommand{\mc}{\mathcal}

\newcommand{\suml}{\sum\limits}
\newcommand{\xra}{\xrightarrow}

{\Large \bf
\begin{center}A hierarchy of behavioral equivalences\\ in the $\pi$-calculus with noisy channels
\end{center}}
\medskip
\centerline{\bf Yongzhi Cao$^{1,2,*}$}
\begin{center}

\medskip
{\small{\it $^1$Institute of Software, School of Electronics
Engineering and Computer Science\\Peking University, Beijing 100871,
China\\$^2$Key Laboratory of High Confidence Software Technologies
(Peking University)\\Ministry of Education, China}

{\it E-mail:} caoyz@pku.edu.cn}
\end{center}

\noindent{\bf Abstract}

{The $\pi$-calculus is a process algebra where agents interact by
sending communication links to each other via noiseless
communication channels. Taking into account the reality of noisy
channels, an extension of the $\pi$-calculus, called the
$\pi_N$-calculus, has been introduced recently. In this paper, we
present an early transitional semantics of the $\pi_N$-calculus,
which is not a directly translated version of the late semantics of
$\pi_N$, and then extend six kinds of behavioral equivalences
consisting of reduction bisimilarity, barbed bisimilarity, barbed
equivalence, barbed congruence, bisimilarity, and full bisimilarity
into the $\pi_N$-calculus. Such behavioral equivalences are cast in
a hierarchy, which is helpful to verify behavioral equivalence of
two agents. In particular, we show that due to the noisy nature of
channels, the coincidence of bisimilarity and barbed equivalence, as
well as the coincidence of full bisimilarity and barbed congruence,
in the $\pi$-calculus does not hold in $\pi_N$.}

{\medskip \noindent{\it Keywords:} $\pi$-calculus, $\pi$-calculus
with noisy channels, barbed equivalence, barbed congruence,
bisimilarity.}

\renewcommand{\thefootnote}{\alph{footnote}}
\setcounter{footnote}{-1} \footnote{ $^*$Supported in part by  the
National Foundation of Natural Sciences of China under Grants
60505011, 60496321, and 60736011.}

\section{Introduction}
$\indent$The need for formal methods in the specification of
concurrent systems has increasingly become well accepted. Particular
interest has been devoted to Petri nets \cite{Pet62,Rei85}, CSP
\cite{Hoa78,Hoa85}, ACP \cite{BerK85}, CCS \cite{Mil80,Mil89}, and
the $\pi$-calculus \cite{MPW92}. The last one due to Milner {\it et
al.} was developed in the late 1980s with the goal of analyzing the
behavior of mobile systems, i.e., systems whose communication
topology can change dynamically, and it turns out to be the unique
one among the aforementioned calculi that can express mobility
directly. The $\pi$-calculus has its roots in CCS, namely CCS with
mobility, introduced by Engberg and Nielsen \cite{EngN86}, while the
capacity of dynamic reconfiguration of logical communication
structure gives the $\pi$-calculus a much greater expressiveness
than CCS.

In the $\pi$-calculus, all distinctions between variables and
constants are removed, communication channels are identified by
names, and computation is represented purely as the communication of
names across channels. The transfer of a name between two agents
(processes) is therefore the fundamental computational step in the
$\pi$-calculus. The basic (monadic) $\pi$-calculus allows only
communication of channel names. There are two extensions of such a
 communication capability: One is the polyadic $\pi$-calculus
\cite{Mil91} that supports communication of tuples, needed to model
passing of complex messages; the other is the higher-order
$\pi$-calculus \cite{San92} that supports the communication of
process abstractions, needed for modeling software composition
within the calculus itself. Interestingly, both of them can be
faithfully translated into the basic $\pi$-calculus.

As we see, communication is a key ingredient of the $\pi$-calculus.
It is worth noting that all communication channels in the
$\pi$-calculus are implicitly presupposed to be noiseless. This
means that in a communication along such a channel, the receiver
will always get exactly what the sender delivers. However, it is
usually not the case in the real world, where communication channels
are often not completely reliable. Recently, Ying \cite{Ying05} took
into account the noise of channels, an idea advocated in his earlier
paper \cite{Ying02b}, and proposed a new version of the
$\pi$-calculus, called the $\pi_N$-calculus. Such a calculus has the
same syntax as that of the $\pi$-calculus. The new feature of
$\pi_N$ arises from a fundamental assumption: Communication channels
in $\pi_N$ may be noisy. This means that what is received at the
receiving end of a channel may be different from what was emitted.

According to a basic idea of Shannon's information theory
\cite{Sha48} that noise can be described in a statistic way,
 the noisy channels in $\pi_N$ was formalized in \cite{Ying05} as follows: Firstly, like
in $\pi$, all (noisy and noiseless) communication channels are
identified by names. Secondly, to describe noise, every pair of
(channel) names $x$ and $y$ is associated with a probability
distribution $p_x (\cdot| y)$ over the output alphabet (here it is
just the set of names), where for any name $z$, $p_x (z | y)$
indicates the probability that $z$ is received from channel $x$ when
$y$ is sent along it. Finally, based on the probability information
arising from the noisy channels, a late probabilistic transitional
semantics of $\pi_N$ is presented. The essential difference between
this semantics and that of $\pi$ is mainly caused by the actions
performed by an output agent $\ol{x}y.P$. In $\pi$, this agent has a
single capability of sending $y$ via channel $x$, expressed as the
transition $\ol{x}y.P\os{\ol{x}y}{\lra} P$, which implies that the
same name $y$ will be received at the receiving end of the channel.
However, because of noise, in the $\pi_N$-calculus the corresponding
transition would be $\ol{x}y.P\os{\ol{x}z}{\lra}_{p_x(z|y)} P.$ This
probabilistic transition indicates that although the name
intentionally sent by the agent is $y$, the name at the receiving
end of the channel $x$ may be the name $z$, different from $y$, with
the probability $p_x (z | y)$. We refer the reader to Section 1.2 in
\cite{Ying05} for a comparison between this model of noisy channels
and the existing literature
\cite{AbdAB99,AbdBRS05,AbdBB01,AbdBJ98,AbdJ96a,AbdJ96b,Ber04,BerH00,HerP00,IyeN97,LuW04,Pri95}
including some works about other formal models with unreliable
communication channels.

It is well known that behavioral equivalences play a very important
role in process algebras because they provide a formal description
that one system implements another. Two agents are deemed equivalent
when they ``have the same behavior" for some suitable notion of
behavior. In terms of the $\pi$-calculus, various behavioral
equivalences have been studied extensively; examples are
\cite{AmaCS98,BorN95,BorS98a,FouG05,MPW92,MilS92,PieS00,San92,San96,SanW01}.
In \cite{Ying05}, some concepts of approximate bisimilarity and
equivalence in CCS \cite{Ying01,Ying02a,Ying03} were generalized
into the $\pi_N$-calculus; such behavioral equivalences involve
quantitative information---probability, since the agent in $\pi_N$
is represented by a probabilistic transition system. To our
knowledge, except for this work there are no probabilistic versions
of behavioral equivalences in the $\pi$-calculus and its variants,
although probabilistic extensions of the $\pi$-calculus
\cite{Her02,HerP00,SylP06} were introduced.

The purpose of this paper is to extend some classical behavioral
equivalences related to (strong) barbed equivalence and (strong)
bisimilarity into $\pi_N$ and cast them in a hierarchy. Following
the model of noisy channels in \cite{Ying05}, we develop an early
transitional semantics of the $\pi_N$-calculus which makes the study
of behavioral equivalences somewhat simpler. It is worthy of note,
however, that unlike in the $\pi$-calculus, this semantics is not a
directly translated version of the late semantics of $\pi_N$ in
\cite{Ying05}. Surprisingly, we have found that not all bound names
in $\pi_N$ are compatible with alpha-conversion when we remove the
strong assumption in \cite{Ying05} that free names and bound names
are distinct. As a result, we have to add a rule for inputting bound
names to the corresponding early semantics of $\pi$. In addition, to
handle transitions of $\pi_N$ well, we group the transitions
according to their sources.

\begin{figure}[htb]\label{FHier}
\begin{minipage}[b]{0.5\textwidth}
\centering \scalebox{.8}{\includegraphics{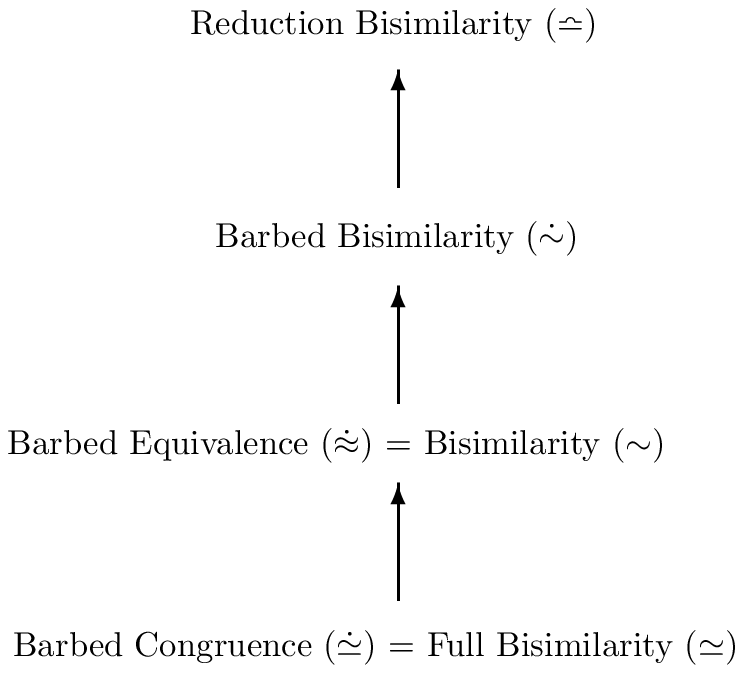}}\vspace{0cm}

\vspace{0.3cm} {\small\bf (a) Hierarchy in the $\pi$-calculus}
\end{minipage}%
\begin{minipage}[b]{0.5\textwidth}
\centering \scalebox{.8}{\includegraphics{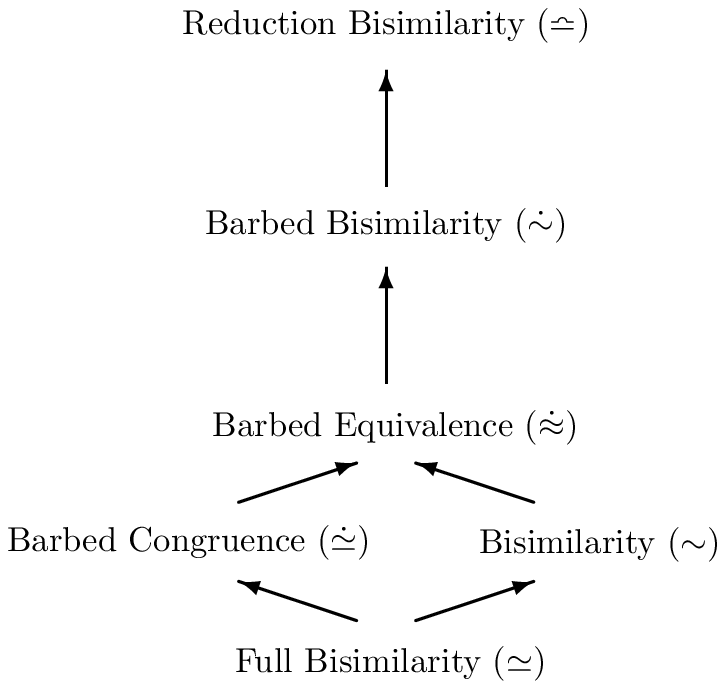}}\vspace{0cm}

{\small\bf (b) Hierarchy in the $\pi_N$-calculus}
\end{minipage}%
\caption{Hierarchy of behavioral equivalences, where an arrow
$A\rightarrow B$ expresses that $A$ is strictly included in $B$}
\end{figure}

Based upon the early transitional semantics of $\pi_N$,  we then
extend six kinds of behavioral equivalences consisting of reduction
bisimilarity, barbed bisimilarity, barbed equivalence, barbed
congruence \cite{MilS92}, bisimilarity, and full bisimilarity
\cite{MPW92,MPW93} into the $\pi_N$-calculus. All these equivalences
are defined through certain bisimulations involving internal action
and discriminating power. Because of noisy channels, a transition
may occur with a certain probability, and thus all the bisimulations
are defined quantitatively. Some basic properties of these
equivalences have been investigated. Finally, we concentrate on a
hierarchy of these behavioral equivalences. In particular, due to
the noisy nature of channels, the coincidence of bisimilarity and
barbed equivalence which holds in the $\pi$-calculus, as well as the
coincidence of full bisimilarity and barbed congruence, fails in the
$\pi_N$-calculus. This hierarchy is shown in Figure \ref{FHier},
together with a corresponding hierarchy in $\pi$ (see, for example,
\cite{SanW01a}). Clearly, the hierarchy is helpful to verify the
behavioral equivalence of two agents: one can start a proof effort
at the middle tier; if this succeeds, one can switch to a finer
equivalence; otherwise, one can switch to a coarser equivalence.

The remainder of this paper is structured as follows. We briefly
review some basics of the $\pi$-calculus in Section 2. After
recalling the formal framework of $\pi_N$ \cite{Ying05} in Section
3.1, we develop the early transitional semantics of the
$\pi_N$-calculus and present it in two different forms in the
remainder of this section. Section 4 is devoted to reduction
bisimilarity and barbed bisimilarity, equivalence, and congruence.
In the subsequent section, the other two equivalences, bisimilarity
and full bisimilarity, are explored. We complete the hierarchy of
these behavioral equivalences in Section 6 and conclude the paper in
Section 7.

\section{$\pi$-calculus}
$\indent$For the convenience of the reader, this section collects
some useful facts on the $\pi$-calculus from \cite{SanW01a}. The
syntax, the early transitional semantics, and some notations of the
$\pi$-calculus are presented in Section 2.1. Section 2.2 briefly
reviews several behavioral equivalences of the $\pi$-calculus. We
refer the
 reader to \cite{mil99,MPW92,par01,SanW01a} and the references therein for an
elaborated explanation and the development of the theory of $\pi$.

\subsection{Basic definitions}
$\indent$We presuppose in the $\pi$-calculus a countably infinite
set ${\bf N}$ of names ranged over by $a, b, \ldots, x, y, \ldots$
with $\tau\not\in{\bf N}$, and such names will act as communication
channels, variables, and data values. We employ $P, Q, R, \ldots$ to
serve as meta-variables of agents or process expressions. Processes
evolve by performing actions, and the capabilities for action are
expressed via the following four kinds of {\it prefixes}:
\begin{eqnarray*}
   \pi&::=& \ol{x}y\;|\;x(z)\;|\;\tau\;|\;[x=y]\pi,
     \end{eqnarray*}
where $\ol{x}y$, an {\it output prefix}, is capable of sending the
name $y$ via the name $x$; $x(z)$, an {\it input prefix}, is capable
of receiving any name via $x$; $\tau$, the {\it silent prefix}, is
an internal action; and $[x=y]\pi$, a {\it match prefix}, has the
capability $\pi$ whenever $x$ and $y$ are the same name.

We now recall the syntax of the $\pi$-calculus.
\begin{Def}\label{DPi}{\rm
  The {\it  processes} and the {\it summations} of the $\pi$-calculus are given respectively by
  \begin{eqnarray*}
   P&::=& M\;|\;P|P'\;|\;(\nu z)P\;|\;!P \\
   M&::=& {\bf0}\;|\;\pi.P\;|\;M+M'.
  \end{eqnarray*}
}
\end{Def}

In the definition above, ${\bf 0}$ is a designated process symbol
that can do nothing. A {\it prefix} $\pi.P$ has a single capability
expressed by $\pi$; the agent $P$ cannot proceed until that
capability has been exercised. A {\it sum} $P+Q$ represents an agent
that can enact either $P$ or $Q$. A {\it parallel composition} $P|Q$
represents the combined behavior of $P$ and $Q$ executing in
parallel, that is, $P$ and $Q$ can act independently, and may also
communicate if one performs an output and the other performs an
input along the same channel. The {\it restriction} operator $(\nu
z)$ in $(\nu z)P$ acts as a static binder for the name $z$ in $P$.
In addition, the input prefix $x(z)$ also binds the name $z$. The
agent $!P$, called {\it replication}, can be thought of as an
infinite composition $P|P|\cdots$ or, equivalently, as a process
satisfying the equation $!P=P|!P$. Iterative or arbitrarily long
behavior can also be described by an alternative mechanism, the
so-called recursion. It turns out that replication can encode the
recursive definition (see, for example, Section 9.5 in
\cite{mil99}).

Let us introduce some syntactic notations before going forward. As
mentioned above, both input prefix and restriction bind names, and
we can define the {\it bound names} bn$(P)$ as those with a bound
occurrence in $P$ and the {\it free names} fn$(P)$ as those with a
not bound occurrence. We write n$(P)$ for the names of $P$, namely,
n$(P)=\mbox{fn}(P)\cup\mbox{bn}(P)$, and sometimes use the
abbreviation fn$(P, Q)$ for fn$(P)\cup\mbox{fn}(Q)$.

A {\it substitution} is a function from names to names that is the
identity except on a finite set. We write $\{y/x\}$ for the
substitution that maps $x$ to $y$ and is identity for all other
names, and in general $\{y_1,\ldots, y_n/x_1,\ldots, x_n\}$, where
the $x_i$'s are pairwise distinct, for a function that maps each
$x_i$ to $y_i$. We use $\sm$ to range over substitutions, and write
$x\sm$, or sometimes $\sm(x)$, for $\sm$ applied to $x$. The process
$P\sm$ is $P$ where all free names $x$ are replaced by $\sm(x)$,
with changes of some bound names (i.e., alpha-conversion) wherever
needed to avoid name captures.


For later need, we fix some notational conventions: A sequence of
distinct restrictions $(\nu z_1)\cdots(\nu z_n)P$ is often
abbreviated to $(\nu z_1\cdots z_n)P$, or just $(\nu
\widetilde{z})P$ when $n$ is not important. We sometimes elide a
trailing ${\bf0}$, writing $\ah$ for the process $\ah.{\bf0}$, where
this cannot cause confusion. We also follow generally used operator
precedence on processes.

For our purpose of investigating barbed equivalence, we only recall
the early transition rules here; the reader may refer to
\cite{MPW92, SanW01a} for the late one. The transition rules are
nothing other than inference rules of labeled transition relations
on processes. The transition relations are labeled by the {\it
actions}, of which there are four kinds: the silent action $\tau$,
input actions $xy$, free output actions $\ol{x}y$, and bound output
actions $\ol{x}(y)$. The first action is internal action, the second
is receiving the name $y$ via the name $x$, the third is sending $y$
via $x$, and the last is sending a fresh name via $x$. Let $\ah,
\beta, \ldots$ range over actions. We write $Act$ for the set of
actions. If $\ah=xy,$ $\ol{x}y,$ or $\ol{x}(y)$, then $x$ is called
the {\it subject} and $y$ is called the {\it object} of $\ah$. The
{\it free names} and {\it bound names} of an action $\ah$ are given
by
\begin{displaymath} \mbox{fn}(\ah)=\left\{
\begin{array}{ll}
\emptyset, & \textrm{if } \ah=\tau\\
\{x, y\}, & \textrm{if } \ah=xy \mbox{ or }\ol{x}y\\
\{x\}, &  \textrm{if } \ah=\ol{x}(y)
\end{array} \right.
\end{displaymath}and
\begin{displaymath} \mbox{bn}(\ah)=\left\{
\begin{array}{ll}
\emptyset, & \textrm{if } \ah=\tau, xy, \mbox{ or }\ol{x}y\\
\{y\}, &  \textrm{if } \ah=\ol{x}(y).
\end{array} \right.
\end{displaymath}The set of names, n$(\ah)$, of $\ah$ is
fn$(\ah)\cup\mbox{bn}(\ah)$.

\begin{table}
{\begin{tabular}{lll} & Out\quad  $\dfrac{\
}{\ol{x}y.P\os{\ol{x}y}{\lra}P}$ & Inp\quad $\dfrac{\ }{x(z).P\os{xy}{\lra}P\{y/z\}}$ \vspace{0.2cm} \\

& Tau\quad $\dfrac{\ }{\tau.P\os{\tau}{\lra}P}$ &
 Mat\quad $\dfrac{\pi.P\os{\ah}{\lra}P'}{[x=x]\pi.P\os{\ah}{\lra}P'}$ \vspace{0.2cm} \\

& Sum-L\quad $\dfrac{P\os{\ah}{\lra}P'}{P+Q\os{\ah}{\lra}P'}$ &
 Par-L\quad $\dfrac{P\os{\ah}{\lra}P'}{P|Q\os{\ah}{\lra}P'|Q}$
\quad bn($\ah$) $\cap$ fn($Q$) $=\emptyset$ \vspace{0.2cm} \\

& Comm-L\quad $\dfrac{P\os{\ol{x}y}{\lra}P'\quad
Q\os{xy}{\lra}Q'}{P|Q\os{\tau}{\lra}P'|Q'}$ & Close-L\quad
$\dfrac{P\xra{\ol{x}(z)}P'\quad
Q\os{xz}{\lra}Q'}{P|Q\os{\tau}{\lra}(\nu z)(P'|Q')}$ \quad $z\not\in$ fn($Q$)\vspace{0.2cm} \\

& Res\quad $\dfrac{P\os{\ah}{\lra}P'}{(\nu z)P\os{\ah}{\lra}(\nu
z)P'}$ \quad $z\not\in$ n($\ah$)&
 Open\quad $\dfrac{P\os{\ol{x}z}{\lra}P'}{(\nu z)P\xra{\ol{x}(z)} P'}$  \quad $z\neq x$\vspace{0.2cm}\\

& Rep-Act\quad $\dfrac{P\os{\ah}{\lra}P'}{!P\os{\ah}{\lra}P'|!P}$ &
Rep-Comm\quad $\dfrac{P\os{\ol{x}y}{\lra}P'\quad
P\os{xy}{\lra}P''}{!P\os{\tau}{\lra}P'|P''|!P}$ \vspace{0.2cm} \\

& Rep-Close\quad $\dfrac{P\xra{\ol{x}(z)}P'\quad
P\os{xz}{\lra}P''}{!P\os{\tau}{\lra}((\nu z)(P'|P''))|!P}$ &
$z\not\in$ fn($P$)\vspace{0.2cm} \\
\end{tabular}}
\caption{Early transition rules of the
$\pi$-calculus\label{Tab:transruleApi}}
\end{table}

The transition relation labeled by $\ah$ will be written as
$\os{\ah}{\lra}$. Thus, $P\os{\tau}{\lra}Q$ will express that $P$
can evolve invisibly to $Q$; $P\os{xy}{\lra}Q$ will express that $P$
can receive $y$ via $x$ and become $Q$; $P\os{\ol{x}y}{\lra}Q$ will
express that $P$ can send $y$ via $x$ and evolve to $Q$; and
$P\os{\ol{x}(y)}{\lra}Q$ will express that $P$ evolves to $Q$ after
sending a fresh name via $x$.

We are now in the position to review the labeled transition
semantics of the $\pi$-calculus. The {\it (early) transition
relations}, $\{\os{\ah}{\lra} \;:\ah\in Act\}$,  are defined by the
rules in Table \ref{Tab:transruleApi}. Note that four rules are
elided from the table: the symmetric forms Sum-R, Par-R, Comm-R, and
Close-R of Sum-L, Par-L, Comm-L, and  Close-L, respectively.

We do not discuss and illustrate the rules, and only remark that the
side conditions in Par-L, Par-R, Close-L, Close-R, Rep-Close can
always be satisfied by changing the object of a bound-output action.

\subsection{Behavioral equivalences}
$\indent$In this subsection, we recall the notions of reduction
bisimilarity, barbed bisimilarity, barbed equivalence, barbed
congruence, bisimilarity, and full bisimilarity of the
$\pi$-calculus studied in \cite{MPW92,MilS92,San92,MPW93,SanW01a}. A
hierarchy of them is also recorded.

Let us begin with reduction bisimilarity.
\begin{Def}
{\rm A relation $\mc{R}$ is a {\it reduction bisimulation} if
  whenever $(P,Q)\in\mc{R}$,

$(1)$ $P\os{\tau}\lra P'$ implies $Q\os{\tau}\lra Q'$ for some $Q'$
with $(P',Q')\in\mc{R}$;

$(2)$ $Q\os{\tau}\lra Q'$ implies $P\os{\tau}\lra P'$ for some $P'$
with $(P',Q')\in\mc{R}$.\\
{\it Reduction bisimilarity}, denoted $\bumpeq$, is the union of all
reduction bisimulations.
  }
\end{Def}

A somewhat stronger notion than reduction bisimilarity but still
very weak is barbed bisimilarity, which takes observability into
account. To express observability formally, we need the notion of
{\it observability predicate} $P\downarrow_\theta$, where $\theta$
is an arbitrary name or co-name, namely $\theta\in{\bf
N}\cup\{\ol{a}:a\in{\bf N}\}$. We say that $P\downarrow_a$ if $P$
can perform an input action with subject $a$; and
$P\downarrow_{\ol{a}}$ if $P$ can perform an output action with
subject $a$.

\begin{Def}
{\rm A relation $\mc{R}$ is a {\it barbed bisimulation} if
  whenever $(P,Q)\in\mc{R}$,

$(1)$ $P\downarrow_\theta$ implies $Q\downarrow_\theta$, and vice
versa;

$(2)$ $P\os{\tau}\lra P'$ implies $Q\os{\tau}\lra Q'$ for some $Q'$
with $(P',Q')\in\mc{R}$;

$(3)$ $Q\os{\tau}\lra Q'$ implies $P\os{\tau}\lra P'$ for some $P'$
with $(P',Q')\in\mc{R}$.\\
{\it Barbed bisimilarity}, written $\dot\sim$, is the union of all
barbed bisimulations.
  }
\end{Def}

Based upon barbed bisimulation, we have the following definition.
\begin{Def}
{\rm Two processes $P$ and $Q$ are called {\it barbed equivalent},
denoted $P\dot\approx Q$, if $P|R\dot\sim Q|R$ for any $R$.
  }
\end{Def}

We are going to introduce barbed congruence, which is stronger than
barbed equivalence. To this end, we need an auxiliary notion.
\begin{Def}{\rm {\it Process contexts} $\mc{C}$ are given by the syntax
\begin{eqnarray*}
   \mc{C}&::=& [\ ]\;|\;\pi.\mc{C}+M\;|\;\mc{C}|P\;|\;P|\mc{C}\;|\;(\nu z)\mc{C}\;|\;!\mc{C}.
  \end{eqnarray*}
We denote by $\mc{C}[P]$ the result of filling the hole $[\ ]$ in
the context $\mc{C}$ with the process $P$. The {\it elementary
contexts} are $\pi.[\ ]+M,\ [\ ]|P,\ P|[\ ],\ (\nu z)[\ ],\
\mbox{and }![\ ]$.}\end{Def}

Now, we can make the following definition.
\begin{Def}
{\rm Two processes $P$ and $Q$ are said to be {\it barbed
congruent}, written $P\dot\simeq Q$, if $\mc{C}[P]\dot\sim
\mc{C}[Q]$ for every process context $\mc{C}$.
  }
\end{Def}

Let us continue to define bisimilarity and an associated congruence,
full bisimilarity.
\begin{Def}
{\rm {\it (Strong) bisimilarity} is the largest symmetric relation,
$\sim$, such that whenever $P\sim Q$, $P\os{\ah}\lra P'$ implies
$Q\os{\ah}\lra Q'$ for some $Q'$ with $P'\sim Q'$.
  }
\end{Def}

\begin{Def}
{\rm Two processes $P$ and $Q$ are {\it full bisimilar}, written
$P\simeq Q$, if $P\sm\sim Q\sm$ for every substitution $\sm$.
  }
\end{Def}

Finally, we summarize a hierarchy of the behavioral equivalences
above, which is depicted in Figure \ref{FHier}(a).
\begin{Thm}\label{ThmChar}{\rm  \

$(1)$
$\dot\simeq\subseteq\dot\approx\subseteq\dot\sim\subseteq\bumpeq$;
each of the inclusions can be strict.

$(2)$ $\sim=\dot\approx$ and $\simeq=\dot\simeq$.
  }
\end{Thm}

For a proof of the above theorem, the reader is referred to Sections
2.1 and 2.2.1 (Lemma 2.2.7 and Theorem 2.2.9) of \cite{SanW01a}. We
 remark that barbed bisimilarity, congruence, and equivalence
were introduced in \cite{MilS92}, the basic theory of bisimilarity
and full bisimilarity was established in \cite{MPW92,MPW93}, and the
assertion $(2)$ of Theorem \ref{ThmChar} was first proved in
\cite{San92}.

\section{$\pi$-calculus with noisy channels}
$\indent$In the last section, we made an implicit assumption that
all communication channels in the $\pi$-calculus are noiseless. In
the present section, such an assumption is removed and the
$\pi$-calculus with noisy channels, the $\pi_N$-calculus, is
explored. The first subsection is devoted to introducing the
original formal framework of the $\pi_N$-calculus due to Ying
\cite{Ying05}, which is based on the late transitional semantics of
$\pi_N$. An early transitional semantics of $\pi_N$ is proposed in
Section 3.2. To cope with transitions of $\pi_N$ well, we group the
transitions according to their sources in Section 3.3.

\subsection{Late transitional semantics of $\pi_N$}
$\indent$This subsection reviews briefly some basic notions of
 the $\pi_N$-calculus from \cite{Ying05}, including noisy channels and the late
transitional semantics.

A fundamental assumption in the $\pi_N$-calculus \cite{Ying05} is
that communication channels may be noisy; that is, their inputs are
subject to certain disturbances in transmission. In other words, the
communication situation is conceived as that an input is transmitted
through a channel and the output is produced at the end of the
channel, but the output is often not completely determined by the
input. In Shannon's information theory \cite{Sha48}, a mathematical
model of channels from statistic communication theory, the noisy
nature is usually described by a probability distribution over the
output alphabet. This distribution of course depends on the input
and in addition it may depend on the internal state of the channel.

A simpler but still very valuable class of noisy channels is
memoryless channels, on which the $\pi_N$-calculus is based. In such
channels, it is assumed that any output does not depend on the
internal state of the channel, and moreover, the outputs of any two
different inputs are independent. It turns out that a memoryless
channel can be completely characterized by its channel matrix
$$[p(y|x)]_{x,y}$$ where $p(y|x)$ is the conditional probability of
outputting $y$ when the input is $x$, and the subscripts $x$ and $y$
run over all inputs and outputs, respectively. Clearly,
$p(y|x)\geq0$, and by definition, we have that $\sum_{y}p(y|x)=1$
for any input $x$.

The syntax of the $\pi_N$-calculus is completely the same as that of
the $\pi$-calculus. The essential difference between $\pi_N$ and
$\pi$ is that $\pi_N$ takes the noise of communication channels into
consideration, which means that the receiver cannot always get
exactly what the sender delivers. As mentioned above, the noisy
channels in $\pi_N$ are assumed to be memoryless, and thus we may
suppose that each name $x\in {\bf N}$ has a channel matrix $$M_x =
[p_x (z | y)]_{y,z\in {\bf N}}$$ where $p_x (z | y)$ is the
probability that the receiver will get the name $z$ at the output
when the sender emits the name $y$ along the channel $x$.

\begin{table}
{\begin{tabular}{llll} {\sc Out} &$\dfrac{\
}{\ol{x}y.P\os{\ol{x}z}{\longmapsto}_pP}$\quad$p=p_x(z|y)>0$ & {\sc
Inp} & $\dfrac{\ }{x(z).P\os{x(z)}{\longmapsto}_1P}$
\vspace{0.2cm} \\

{\sc Tau} & $\dfrac{\ }{\tau.P\os{\tau}{\longmapsto}_1P}$ &
{\sc  Mat} & $\dfrac{\pi.P\os{\ah}{\longmapsto}_pP'}{[x=x]\pi.P\os{\ah}{\longmapsto}_pP'}$ \vspace{0.2cm} \\

{\sc Sum-L} &
$\dfrac{P\os{\ah}{\longmapsto}_pP'}{P+Q\os{\ah}{\longmapsto}_pP'}$
&\ &\ \vspace{0.2cm} \\

{\sc Par-L}&
$\dfrac{P\os{\ah}{\longmapsto}_pP'}{P|Q\os{\ah}{\longmapsto}_pP'|Q}$
 &\hspace{-2.8cm} bn$(\ah)\cap\mbox{fn}(Q)=\emptyset$ &\
  \vspace{0.2cm} \\

{\sc Comm-L} & $\dfrac{P\os{\ol{x}y}{\longmapsto}_pP'\quad
Q\os{x(z)}{\longmapsto}_1Q'}{P|Q\os{\tau}{\longmapsto}_pP'|Q'\{y/z\}}$
& {\sc Close-L} & $\dfrac{P\os{\ol{x}(z)}\longmapsto_pP'\quad
Q\os{x(z)}{\longmapsto}_1Q'}{P|Q\os{\tau}{\longmapsto}_p(\nu z)(P'|Q')}$  \vspace{0.2cm} \\

{\sc Open} & $\dfrac{P\os{\ol{x}y}{\longmapsto}_pP'}{(\nu
y)P\os{\ol{x}(y)}\longmapsto_p P'}$&\  &\hspace{-4.0cm}$y\neq
x$\vspace{0.2cm}\\

{\sc Res} & $\dfrac{P\os{\ah}{\longmapsto}_pP'}{(\nu
z)P\os{\ah}{\longmapsto}_p(\nu z)P'}$ \quad $z\not\in$ n($\ah$)&
{\sc Rep-Act} &
$\dfrac{P\os{\ah}\longmapsto_{p}P'}{!P\os{\ah}\longmapsto_{p}P'|!P}$\vspace{0.2cm}\\

{\sc Rep-Comm} & $\dfrac{P\os{\ol{x}y}\longmapsto_{p}P'\quad
P\os{x(z)}{\longmapsto}_1P''}{!P\os{\tau}{\longmapsto}_{p}P'|P''\{y/z\}|!P}$&
{\sc Rep-Close} & $\dfrac{P\os{\ol{x}(z)}\longmapsto_{p}P'\quad
P\os{x(z)}{\longmapsto}_1P''}{!P\os{\tau}{\longmapsto}_{p}(\nu
z)(P'|P'')|!P}$\vspace{0.2cm}
\end{tabular}}
\caption{Late transition rules of $\pi_N$\label{T:latepiNtr}}
\end{table}

For the $\pi$-calculus, there are two kinds of transitional
semantics. In Section 2.1, the input rule is a transition
$x(z).P\os{xy}{\lra}P\{y/z\}$, which expresses that $x(z).P$ can
receive the name $y$ via $x$ and evolve to $P\{y/z\}$. An action of
the form $xy$ records both the name used for receiving and the name
received. The placeholder $z$ is instantiated early, namely when the
input by the receiver is inferred. Hence the name ``early"
semantics. In the literature on $\pi$, the first way to treat the
semantics for input, the late transitional semantics \cite{MPW92},
adopts the input rule $x(z).P\os{x(z)}{\longmapsto}P,$ where the
label $x(z)$ contains a placeholder $z$ for the name to be received,
rather than the name itself. In this context, the input action of
the form $x(z)$ which replaces the action $xy$ in the early
transitional semantics can be instantiated late, that is, it can be
instantiated when a communication is inferred. Therefore, the early
and late terminology is based upon when a name (placeholder) is
instantiated in inferring an interaction. In fact, there is a very
close relationship between the two kinds of semantics (see, for
example, Lemma 4.3.2 in \cite{SanW01a}) which allows us to freely
use the early or late semantics as convenient.

Let us write $Act_l$ for the set of actions in the late transitional
semantics, namely, $Act_l=\{\tau, x(y), \ol{x}y, \ol{x}{(y)}: x,y\in
\bf{N}\}$. If $\ah=x(y)$, we set fn$(\ah)=\{x\}$ and
bn$(\ah)=\{y\}$. The structural operational semantics of $\pi_N$ in
\cite{Ying05} is based upon the late transitional semantics and is
given by a family of probabilistic transition relations
$\os{\ah}{\longmapsto}_p$ $(\ah\in Act_l, p\in(0,1])$ displayed in
Table \ref{T:latepiNtr}. The table omits the symmetric forms of {\sc
Sum-L, Par-L, Comm-L,} and {\sc Close-L}; note also that the rule
IDE for agent identifiers in \cite{Ying05} is replaced by the rules
{\sc Rep-Act, Rep-Comm,} and {\sc Rep-Close} since we are using an
equivalent notion, replications, instead of agent identifiers in the
syntax. The arrow $\longmapsto$ is used to distinguish the late
relations from the early, and the probability values $p$ arise
entirely from the noise of communication channels. The {\sc Out}
rule, which represents the noisy nature of channels, is the unique
one that all differences between $\pi_N$ and $\pi$ come from. It
means that the process $\ol{x}y.P$ of output prefix form sends the
name $y$ via the channel $x$, but what the receiver gets at the
output of this channel may not be $y$ due to noise residing in it,
and a name $z$ will be received with the probability $p_x (z | y)$.

\subsection{Early transitional semantics of $\pi_N$}
$\indent$For the $\pi$-calculus, it turns out that the early
transitional semantics is somewhat simpler than the late one for
investigating behavioral equivalences. The reason is that the input
action in the early semantics is instantiated early and thus we need
not check all possible instantiations of a placeholder. In light of
this, we pay our attention to the early transitional semantics of
$\pi_N$ in this subsection. This semantics is not, however, a direct
translation of Table \ref{T:latepiNtr}, as we will see shortly.

\begin{table}\centering
{\begin{tabular}{lllllll}$\ah$ & kind & barb($\ah$)  & subj($\ah$)&
obj($\ah$) & n($\ah$) & $\ah\sigma$
\\ \cline{1-7} $\tau$ & Silent &
\multicolumn{1}{c}{$\tau$}& \multicolumn{1}{c}{$-$} &
\multicolumn{1}{c}{$-$} &
\multicolumn{1}{c}{$\emptyset$}  & $\tau$\\
$xy$ & Input & \multicolumn{1}{c}{$x$}& \multicolumn{1}{c}{$x$}&
\multicolumn{1}{c}{$y$} & \multicolumn{1}{c}{$\{x,y\}$}
  & $x\sigma y\sigma$\\
$\ol{x}y$ & Noisy free output & \multicolumn{1}{c}{$\ol{x}$}&
\multicolumn{1}{c}{$x$} & \multicolumn{1}{c}{$y$} &
\multicolumn{1}{c}{$\{x,y\}$}  &
$\ol{x\sigma}y\sigma$\\
$\ol{x}(y)$ & Noisy bound output & \multicolumn{1}{c}{$\ol{x}$}&
\multicolumn{1}{c}{$x$} & \multicolumn{1}{c}{$y$} &
\multicolumn{1}{c}{$\{x,y\}$}  & $\ol{x\sigma}(y\sigma)$
\end{tabular}}
\caption{Terminology and notation for actions\label{Tab:names}}
\end{table}

Like the late transitional semantics of $\pi_N$, the early semantics
of $\pi_N$ is also given in terms of probabilistic transition
relations. A probabilistic transition in the $\pi_N$ is of the form
$$P{\os{\ah}\lra}_{p}Q$$ where $P$ and $Q$ are two processes,
$\ah\in Act=\{\tau, xy, \ol{x}y, \ol{x}{(y)}: x,y\in \bf{N}\}$, and
$p\in(0,1]$. The intuitive meaning of this transition is that the
agent $P$ performs action $\ah$ and becomes $Q$, with probability
$p$. It should be pointed out that although the actions here are the
same as those in the early semantics of $\pi$, the meanings of them
are not completely identical. More concretely, $\tau$ and $xy$ still
represent an internal action and an input of a name $y$ alone
channel $x$, respectively, but $\ol{x}y$ represents output of a name
$y$ via a noisy channel $x$ that changes the intended output of some
name into $y$ with a certain probability, and $\ol{x}{(y)}$
represents output of a bound name $y$ via a noisy channel $x$ that
changes the intended output of some name into $y$ with a certain
probability. Table \ref{Tab:names} displays terminology and notation
pertaining to the actions. Its columns list, respectively, the {\it
kind} of an action $\ah$, the {\it barb} of $\ah$, the {\it subject}
of $\ah$, the {\it object} of $\ah$, the set of {\it names} of
$\ah$, and the effect of applying a substitution to $\ah$; some
issues different from those of $\pi$ such as $\ol{x}(y)\sm$ will be
explained subsequently.

\begin{table}
{\begin{tabular}{llll} Out & $\dfrac{\
}{\ol{x}y.P\os{\ol{x}z}{\lra}_pP}$\quad $p=p_x(z|y)>0$\quad\qquad &
Inp & $\dfrac{\ }{x(z).P\os{xy}{\lra}_1P\{y/z\}}$
\vspace{0.2cm} \\

Tau & $\dfrac{\ }{\tau.P\os{\tau}{\lra}_1P}$ &
 Mat & $\dfrac{\pi.P\os{\ah}{\lra}_pP'}{[x=x]\pi.P\os{\ah}{\lra}_pP'}$ \vspace{0.2cm} \\

Sum-L & $\dfrac{P\os{\ah}{\lra}_pP'}{P+Q\os{\ah}{\lra}_pP'}$ & Par-L
& $\dfrac{P\os{\ah}{\lra}_pP'}{P|Q\os{\ah}{\lra}_pP'|Q}$
  \vspace{0.2cm} \\

Comm-L & $\dfrac{P\os{\ol{x}y}{\lra}_pP'\quad
Q\os{xy}{\lra}_1Q'}{P|Q\os{\tau}{\lra}_pP'|Q'}$ & Close-L &
$\dfrac{P\xra{\ol{x}(y)}_pP'\quad
Q\os{xy}{\lra}_1Q'}{P|Q\os{\tau}{\lra}_p(\nu y)(P'|Q')}$  \vspace{0.2cm} \\

Res & $\dfrac{P\os{\ah}{\lra}_pP'}{(\nu z)P\os{\ah}{\lra}_p(\nu
z)P'}$  \quad $z\not\in$ n($\ah$)& Open-Out &
$\dfrac{P\os{\ol{x}y}{\lra}_pP'}{(\nu y)P\xra{\ol{x}(y)}_p P'}$
\quad $y\neq x$\vspace{0.2cm}\\

Open-Inp & $\dfrac{P\os{xy}{\lra}_1P'}{(\nu y)P\os{xy}{\lra}_1P'}$
\quad $y\neq x$& Rep-Act &
$\dfrac{P\os{\ah}\lra_{p}P'}{!P\os{\ah}\lra_{p}P'|!P}$\vspace{0.2cm}\\

Rep-Comm & $\dfrac{P\xra{\ol{x}y}_{p}P'\quad
P\os{xy}{\lra}_1P''}{!P\os{\tau}{\lra}_{p}P'|P''|!P}$& Rep-Close &
$\dfrac{P\xra{\ol{x}(y)}_{p}P'\quad
P\os{xy}{\lra}_1P''}{!P\os{\tau}{\lra}_{p}(\nu y)(P'|P'')|!P}$
\end{tabular}}\vspace{.2cm}
\caption{Early transition rules of $\pi_N$\label{T:piNtr}}
\end{table}

The early transition rules of $\pi_N$ is present in Table
\ref{T:piNtr}. As before, we omit the symmetric forms of Sum-L,
Par-L, Comm-L, and Close-L. Let us make a brief discussion about the
rationale behind the design:

1) Since we follow the assumption of the noisy channels in
\cite{Ying05}, the Out rule is the same as {\sc Out} in Table
\ref{T:latepiNtr}. It shows that the action performed by $\ol{x}y.P$
is not $\ol{x}y$ but $\ol{x}z$, and the probability $p_x (z | y)$
that $y$ becomes $z$ in channel $x$ is indicated. This is thought of
as that noise happens at the end of sending, not at the end of
receiving.

2) All rules except for Out and Open-Inp are just simple imitations
of the corresponding rules in the $\pi$-calculus. Nevertheless,
there are two differences: one is that a probability parameter $p$
is taken into account, which is necessary for encoding the noise of
channels; the other is that the side conditions in Par-L, Close-L,
and Rep-Close are elided. The latter arises entirely from that the
condition bn$(\ah)\cap\mbox{fn}(Q)=\emptyset$ is not required when
considering the left parallel composition $P|Q$.

The reason for removing  the condition
bn$(\ah)\cap\mbox{fn}(Q)=\emptyset$ arises from the following
consideration. Recall that in $\pi_N$ \cite{Ying05} it was supposed
that free names and bound names are distinct. This assumption
inconveniences the use of some transition rules such as Open-Out in
our context. Recall also that in $\pi$ the side condition
bn$(\ah)\cap\mbox{fn}(Q)=\emptyset$ of inferring $P|Q$ can be easily
satisfied by utilizing alpha-conversion on $P$. However, this
conversion must involve the congruent equation $(\nu z)P\equiv(\nu
w)P\{w/z\}$, where $w\not\in \mbox{n}(P)$. It is unfortunate that
such a well known and widely used equation in the concurrency
community seems to be impracticable for $\pi_N$. To see this, let us
examine a specific example. Suppose that $$p_x(y|y) =p_0,\quad
p_x(a|y)=p_1,\quad p_x(b|y)=p_2,$$ where $p_0+p_1+p_2=1$.We choose
$P\os{\rm def}{=}(\nu a)\ol{x}y|x(w).\ol{w}z$ and $Q\os{\rm
def}{=}(\nu b)\ol{x}y|x(w).\ol{w}z$. Assume that $(\nu z)P\equiv(\nu
w)P\{w/z\}$ with $w\not\in \mbox{n}(P)$ holds in $\pi_N$. Then it is
clear that $P\equiv Q$, and thus we can identify $P$ with $Q$. By
the early transition rules of $\pi_N$, we get without breaking the
side condition bn$(\ah)\cap\mbox{fn}(Q)=\emptyset$ that
\begin{align*}
P\os{\tau}{\lra}_{p_0}\ol{y}z,\ &\ P\os{\tau}{\lra}_{p_1}(\nu
a)\ol{a}z,\ \ P\os{\tau}{\lra}_{p_2}\ol{b}z; \\
Q\os{\tau}{\lra}_{p_0}\ol{y}z,\ &\ Q\os{\tau}{\lra}_{p_1}\ol{a}z,\ \
Q\os{\tau}{\lra}_{p_2}(\nu b)\ol{b}z.
\end{align*} Because $(\nu
a)\ol{a}z$ and $(\nu b)\ol{b}z$ are inactive, while $\ol{x}z$  and
$\ol{b}z$ are capable of sending $z$, this forces that $p_1=p_2=0$,
and thus $p_0=1$. It means that the channel $x$ is noiseless when
outputting $y$; this is absurd because $x$ and $y$ can be taken
arbitrarily, including noisy channels.

In light of the previous discussion, it seems better to do away with
the congruent equation $(\nu z)P\equiv(\nu w)P\{w/z\}$, $w\not\in
\mbox{n}(P)$. As a result, we cannot keep the side condition
bn$(\ah)\cap\mbox{fn}(Q)=\emptyset$, because otherwise the
associative law among process interactions would be violated. To see
this, one may consider the processes $x(z)|((\nu y)\ol{x}y|\ol{y}w)$
and $(x(z)|(\nu y)\ol{x}y)|\ol{y}w$. Since the associativity is a
very important property of mobile systems that the $\pi$-calculus
has intended to understand, we would not like to destroy it. As a
consequence, the side conditions in Par-L, Close-L, and Rep-Close
are removed, and this yields that the proposed early transitional
semantics of $\pi_N$ is somewhat different from that of $\pi$ when
considering only noiseless channels.

Recall also that in the $\pi$-calculus, a private name in a process
$P$ is local, meaning it can be used only for communication between
components within $P$. Such a private name cannot immediately be
used as a port for communication between $P$ and its environment; in
fact, because $P$ may rename its private names, these names are not
known by the environment. Note, however, that not allowing the
congruence $(\nu z)P\equiv(\nu w)P\{w/z\}$, together with the noise
of channels, makes the private names in $\pi_N$ somewhat public.
More concretely, since a process $P$ in $\pi_N$ cannot rename its
private names, these names may appear in the environment of $P$. In
addition, every name may be confused with any other name because of
the noise of channels. Nevertheless, private names are needed in
$\pi_N$ since a private name, say $z$, can at least be used to
preclude a process from communicating with its environment via the
port $z$.

3) An Open-Inp rule is added. This arises from two aspects of
consideration: One is that if a channel is capable of inputting,
then it should have the ability of arbitrary inputting. In other
words, if $P\os{xy}{\lra}_{1}Q$, then restricting $y$ to $P$ should
not prevent the channel $x$ from inputting $y$. The other aspect is
that if an agent can receive a name from outside, then the name may
be thought of as open and the scope of the restriction may be
extended. Technically, the Open-Inp rule derives from the
invalidation of the congruent equation $(\nu z)P\equiv(\nu
w)P\{w/z\}$, $w\not\in \mbox{n}(P)$, because without the equation,
 bound names cannot be changed and the free name in an input may clash with bound names.
We remark that such a rule is not necessary in the $\pi$-calculus,
since all the bound names in $\pi$ can be renamed by
alpha-conversion.

Let us continue introducing some notions. As in $\pi$, the input
prefix $x(z)$ and the restriction $(\nu z)$ bind the name $z$. In
view of 2) above, in the $\pi_N$-calculus we need to differentiate
between the bound names in $(\nu z).Q$ and $x(z).Q$. A bound name
$z$ is called {\it strongly bound} in $P$ if it lies within some
sub-term $x(z).Q$ of $P$. As shown in
 Definition \ref{DStrCong} $(1)$, we permit of changing a strongly
bound name into a fresh name, which is called {\it strong
alpha-conversion}. Any bound name that is not strongly bound is said
to be {\it weakly bound}. An occurrence of a name in a process of
$\pi_N$ is {\it free} if it is not bound. For instance, in
$P\os{{\rm def}}{=}(\nu s)x(z).(\nu z)\ol{y}z.\ol{x}s$, the name $s$
is weakly bound, $z$ is strongly bound, and $x$ and $y$ are free. We
denote the free names, bound names, strongly bound names, and weakly
bound names in a process $P$ by fn$(P)$, bn$(P)$, sbn$(P)$, and
wbn$(P)$, respectively.

We now define the effect of applying a substitution $\sm$ to a
process $P$ in $\pi_N$. Since weakly bound names cannot be
converted, the process $P\sm$ is $P$ where all free names and weakly
bound names $x$ are replaced by $x\sm$, with strong alpha-conversion
wherever needed to avoid captures. This means that strongly bound
names are changed such that whenever $x$ is replaced by $x\sm$ then
the so obtained occurrence of $x\sm$ is not strongly bound. For
instance,
$$(a(x).(\nu b)\ol{x}b.\ol{c}y.{\bf0})\{x,c/y,b\}=a(z).(\nu c)\ol{z}c.\ol{c}x.{\bf0}.$$

Clearly, according to the above definition of substitution, we have
the following fact.
\begin{Lem}{\rm
  For any substitution $\sm$,
\begin{enumerate}
\item[(1)] ${\bf 0}\sm={\bf 0}$;
\item[(2)] $(\pi.P)\sm=\pi\sm.P\sm$;
\item[(3)] $(P+Q)\sm=P\sm+Q\sm$;
\item[(4)] $(P|Q)\sm=P\sm|Q\sm$;
\item[(5)] $((\nu z)P)\sm=(\nu z\sm)P\sm$;
\item[(6)] $(!P)\sm=!P\sm$.
\end{enumerate}
}\end{Lem}

Since the notion of free names is not sufficient for defining
structural congruence, we introduce an extended notion of free
names. Motivated by a similar notion in \cite{Ying05}, we define the
set of noisy free names to be the set of all free names in a process
and those names produced by noise when sending free names. Formally,
we have the following.

\begin{Def}{\rm The set of {\it noisy free names}, denoted
fn$^*(P)$, is defined inductively as follows:
\begin{enumerate}
    \item[(1)] $\mbox{fn}^{*}({\bf 0})=\emptyset$;
    \item[(2)] $\mbox{fn}^{*}(\ol{x}y.P)=\{x\}\cup\{z\in{\bf N}: p_x(z|y)>0\}\cup\mbox{fn}^{*}(P)$;
    \item[(3)] $\mbox{fn}^{*}(x(z).P)=\{x\}\cup\bigcup_{y\in\bf{N}}(\mbox{fn}^{*}(P\{y/z\})\backslash\{y\})$;
    \item[(4)] $\mbox{fn}^{*}(\tau.P)=\mbox{fn}^{*}(P)$;
    \item[(5)] $\mbox{fn}^{*}([x=y]\pi.P)=\{x,y\}\cup\mbox{fn}^{*}(\pi.P)$;
    \item[(6)] $\mbox{fn}^{*}(P+P')=\mbox{fn}^{*}(P|P')=\mbox{fn}^{*}(P)\cup\mbox{fn}^{*}(P')$;
    \item[(7)] $\mbox{fn}^{*}((\nu z)P)=\mbox{fn}^{*}(P)\backslash\{z\}$;
    \item[(8)] $\mbox{fn}^{*}(!P)=\mbox{fn}^{*}(P)$.
\end{enumerate}
}
\end{Def}

Although fn$^*(P)$ is an extension of fn$(P)$, it is not necessarily
that fn$^*(P)\supseteq\mbox{fn}(P)$. For example, assume that
$p_x(z|y)=1$. Then we see that fn$^*(\ol{x}y.{\bf0})=\{x,z\}$, while
fn$(\ol{x}y.{\bf0})=\{x,y\}$. If it is required that $p_x(y|y)>0$
for all $x,y\in{\bf N}$, then we indeed have that
fn$^*(P)\supseteq\mbox{fn}(P)$.

We can now define structural congruence as follows.
\begin{Def}\label{DStrCong}{\rm
Two process expressions $P$ and $Q$ in the $\pi_N$-calculus are {\it
structurally congruent}, denoted $P\equiv Q$,
  if we can transform one into the other by using the following equations (in either direction):
\begin{enumerate}
\item[(1)] $x(z).P\equiv x(w).P\{w/z\}$ if $w$ is fresh in $P$.
\item[(2)] Reordering of terms in a summation.
\item[(3)] $M+{\bf0}\equiv M,\ P|{\bf0}\equiv P,\ P|Q\equiv Q|P,\mbox{ and } P|(Q|R)\equiv (P|Q)|R.$
\item[(4)] $(\nu z)(P|Q)\equiv P|(\nu z)Q$ if $z\not\in$ fn$^*$($P$), $(\nu z){\bf 0}\equiv {\bf 0},
    \mbox{ and } (\nu z)(\nu w)P\equiv (\nu w)(\nu z)P$.
\item[(5)] $[x=x]\pi.P\equiv\pi.P.$
\item[(6)] $[x=y]\pi.P\equiv{\bf 0}$ if $x$ and $y$ are distinct.
\item[(7)] $!P\equiv P|!P.$
\end{enumerate}}
  \end{Def}

The noisy free names of two structurally congruent processes are
clearly related by the following fact.
\begin{Lem}{\rm
If $P\equiv Q$ can be inferred without using $(5)$ and $(6)$ in
Definition \ref{DStrCong}, then $\mbox{fn}^*(P)=\mbox{fn}^*(Q)$.
}\end{Lem}

We also make an observation on the noisy free names of processes
appearing in the same transition.
\begin{Lem}{\rm
  Let $P\os{\ah}{\lra}_pP'$ be a transition in $\pi_N$.
  \begin{enumerate}
    \item[(1)] If $\ah=\ol{x}y$, then $x,y\in\mbox{fn}^*(P)$ and
    $\mbox{fn}^*(P')\subseteq\mbox{fn}^*(P)$.
    \item[(2)] If $\ah=xy$, then $x\in\mbox{fn}^*(P)$ and
    $\mbox{fn}^*(P')\subseteq\mbox{fn}^*(P,y)$.
    \item[(3)] If $\ah=\ol{x}(y)$, then $x\in\mbox{fn}^*(P)$ and
    $\mbox{fn}^*(P')\subseteq\mbox{fn}^*(P,y)$.
    \item[(4)] If $\ah=\tau$, then $\mbox{fn}^*(P')\subseteq\mbox{fn}^*(P)$.
  \end{enumerate}
}\end{Lem}
\begin{proof}The proof is carried out by induction on the depth of
inference $P\os{\ah}{\lra}_pP'$. We need to consider all kinds of
transition rules that are possible as the last rule in deriving
$P\os{\ah}{\lra}_pP'$. The assertions $(1)$ and $(2)$ can be
verified directly, the proof of $(3)$ needs $(1)$, and the proof of
$(4)$ needs the first three. This is a long but routine case
analysis, so the details are omitted.
\end{proof}

We end this subsection by providing some image-finiteness properties
of probabilistic transitions. To this end, we suppose that
outputting a name can only gives rise to finite noisy names, that
is, we adopt the following convention:
\begin{Conv}\label{Conv}{\rm
  For any $x,y\in{\bf N}$, the set $\{y_i: p_x(y_i|y)>0\}$
is finite. }\end{Conv}

The following facts about input and output actions are the
corresponding results of Lemmas 1.4.4 and 1.4.5 in \cite{SanW01a}.
\begin{Lem}\label{Lfo}{\rm \

$(1)$ If $P\os{xy}{\lra}_1P'$ and
$y\not\in\mbox{fn}(P)\cup\mbox{wbn}(P)$, then $P\os{xz}{\lra}_1
P'\{z/y\}$ for any $z$.

$(2)$ If $P\os{xy}{\lra}_1P'$ and
$z\not\in\mbox{fn}(P)\cup\mbox{wbn}(P)$, then there is $P''$ such
that $P\os{xz}{\lra}_1 P''$ and $P''\{y/z\}=P'$.

$(3)$ For any $P$ and $x$, there exist $P_1,\ldots,P_n$ and
$y\not\in\mbox{fn}(P)\cup\mbox{wbn}(P)$ such that if
$P\os{xz}{\lra}_1 P'$ then $P'=P_i\{z/y\}$ for some $P_i$.

$(4)$ For any $P$, there are only finitely many $x$ such that
$P\os{xy}{\lra}_1 P'$ for some $y$ and $P'$.
   }\end{Lem}
\begin{proof}
The first two assertions can be easily proved by induction on
inference, and the last two by induction on $P$.
\end{proof}

Similar to the above, we have a result about output actions.
\begin{Lem}\label{Lbo}{\rm \

$(1)$ For any $P$, there are only finitely many quadruples $x, y, p,
P'$ such that $P\os{\ol{x}y}{\lra}_{p}P'$.

$(2)$ For any $P$, there are only finitely many quadruples $x, y, p,
P'$ such that $P\os{\ol{x}(y)}{\lra}_{p}P'$.
   }\end{Lem}
\begin{proof}All the two assertions are proved by induction on $P$.
The proof of $(1)$ needs Convention \ref{Conv}, and the proof of
$(2)$ uses the fact that weakly bound names cannot be converted.
\end{proof}

\subsection{Transition groups of $\pi_N$}
$\indent$In the last subsection, the primitive transition rules of
$\pi_N$ have been established. A closer examination, however, shows
that at least two confusions may arise from the presentation of
these rules. To avoid this, we introduce another presentation,
transition groups, in this subsection.

To illustrate our motivation, let us consider a simple example: Take
$P\os{{\rm def}}{=}\ol{x}y+\ol{x}z$, and suppose that
\begin{equation*}
\begin{split}
p_x(y|y)=0.7,\ p_x(z|y)=0.1,\ p_x(s|y)=0.1,\ p_x(t|y)=0.1;\\
p_x(y|z)=0.5,\ p_x(z|z)=0.3,\ p_x(s|z)=0.1,\ p_x(w|z)=0.1.
\end{split}
\end{equation*}
By the transition rules of $\pi_N$, we see that
\begin{equation*}
\begin{split}
P\os{\ol{x}y}{\lra}_{0.7}{\bf0},\ P\os{\ol{x}y}{\lra}_{0.5}{\bf0},\
P\os{\ol{x}s}{\lra}_{0.1}{\bf0},\mbox{ and }
P\os{\ol{x}t}{\lra}_{0.1}{\bf0}.
\end{split}
\end{equation*}

At this point, two confusions arise: One is that there are two
probabilistic transitions $P\os{\ol{x}y}{\lra}_{0.7}{\bf0}$ and
$P\os{\ol{x}y}{\lra}_{0.5}{\bf0}$ with the same source and target
processes and the same action, but different probability values. In
\cite{Ying05}, it is thought that the probability of transition
$P\os{\ol{x}y}{\lra}{\bf0}$ is either $0.7$ or $0.5$, but which of
them is not exactly known and the choice between $0.7$ and $0.5$ is
made by the environment. This understanding that comes from the idea
of imprecise probability studied widely by the communities of
Statistics and Artificial Intelligence (see, for example,
\cite{wal91}) is very natural. The limit is that more uncertainties
are involved in inference. Another way to deal with this problem in
the literature on probabilistic processes is to modify probabilistic
transition systems. This is done by adding up all possible values of
transition probability with the same source and target agents and
the same action, and then normalizing them if necessary (for
instance, see \cite{GlaSST90,LarS91,GlaSS95,Ying02b}). Once again,
this kind of modification highly complicates the theory of
probabilistic processes.

The other confusion is that the information on the origins of some
probabilistic transitions is lost. As for the above example, the
information that $P\os{\ol{x}t}{\lra}_{0.1}{\bf0}$ derives
 from the first sub-term of $P$ is lost. In addition, we cannot
determine from which the transition
$P\os{\ol{x}s}{\lra}_{0.1}{\bf0}$ arises.

There is a convenient presentation for alleviating the confusions.
In fact, we may group all transitions having the same origin.
Regarding the example above, we may say that $P$ has two transition
groups
\begin{equation*}
\begin{split}
P\{{\os{\ol{x}y}\lra}_{0.7}{\bf0}, {\os{\ol{x}z}\lra}_{0.1}{\bf0},
{\os{\ol{x}s}\lra}_{0.1}{\bf0}, {\os{\ol{x}t}\lra}_{0.1}{\bf0}\}
\mbox{ and }P\{{\os{\ol{x}y}\lra}_{0.5}{\bf0},
{\os{\ol{x}z}\lra}_{0.3}{\bf0}, {\os{\ol{x}s}\lra}_{0.1}{\bf0},
{\os{\ol{x}w}\lra}_{0.1}{\bf0}\}.
\end{split}
\end{equation*}Notice that the agent $P\os{{\rm def}}{=}\ol{x}y+\ol{x}z$ has a
nondeterministic choice between $\ol{x}y$ and $\ol{x}z$, and it is
usually thought that such a choice is made by the environment, so we
may think that the choice between the transition groups is also made
by the environment. In this way, the first confusion is completely
excluded, and if we have observed an output $\ol{x}t$ and the
environment does not change her choice, then the process has a
smaller probability of outputting $\ol{x}z$ and the output $\ol{x}s$
arises necessarily from the first sub-term of $P$.

In general, we use $$P\{{\os{\ah_i}\lra}_{p_i}Q_i\}_{i\in I}$$ to
represent a group of probabilistic transitions
$P{\os{\ah_i}\lra}_{p_i}Q_i$, ${i\in I}$, satisfying $\sum_{i\in
I}p_i=1$; $P\{{\os{\ah_i}\lra}_{p_i}Q_i\}_{i\in I}$ is called a {\it
transition group}. We omit the indexing set $I$ whenever $I$ is a
singleton. In fact, some representations analogous to the transition
group have already been used in the literature (see, for example,
\cite{SegL95,HerP00,SylP06,DenP07}).

To manipulate transition groups, we need the operator $\uplus$
defined as follows:
\begin{displaymath} \{{\os{\ah_i}\lra}_{p_i}Q_i\}_{i\in I}\uplus\{{\os{\beta}\lra}_{p}Q\}=\left\{
\begin{array}{ll}
\{{\os{\ah_i}\lra}_{p_i+p}Q_i\}_{i\in I}, & \textrm{if $Q=Q_i$ and $\beta=\ah_i$ for some $i$}\\
\{{\os{\ah_i}\lra}_{p_i}Q_i\}_{i\in I}\cup\{{\os{\beta}\lra}_{p}Q\},
& \textrm{otherwise};
\end{array} \right.
\end{displaymath}
\begin{displaymath}
\{{\os{\ah_i}\lra}_{p_i}Q_i\}_{i\in
I}\uplus\{{\os{\beta_j}\lra}_{p_j}Q_j\}_{j\in
J}=\big(\{{\os{\ah_i}\lra}_{p_i}Q_i\}_{i\in
I}\uplus\{{\os{\beta_j}\lra}_{p_j}Q_j\}\big)\uplus\{{\os{\beta_k}\lra}_{p_k}Q_k\}_{k\in
J\backslash\{j\}}.\qquad\;
\end{displaymath}
Moveover, if in the same transition group, there are
$P{\os{\ah_i}\lra}_{p_i}Q_i$ and $P{\os{\ah_j}\lra}_{p_j}Q_j$ with
$\ah_i=\ah_j$ and $Q_i=Q_j$, then we sometimes combine them into a
single one $P{\os{\ah_i}\lra}_{p_i+p_j}Q_i$ and delete $j$ from the
indexing set $I$. For instance, the transition groups
$P\{{\os{\ol{x}y}\lra}_{0.4}Q\}\uplus\{{\os{\ol{x}y}\lra}_{0.6}Q\}$,
$P\{{\os{\ol{x}y}\lra}_{0.4}Q, {\os{\ol{x}y}\lra}_{0.6}Q\}$, and
$P\{{\os{\ol{x}y}\lra}_{1}Q\}$ mean the same thing.

The following result is helpful to group the transitions of $\pi_N$.
\begin{Lem}\label{Lsa}{\rm Let $P\{{\os{\ah_i}\lra}_{p_i}Q_i\}_{i\in I}$ be a
transition group derived from the early transition rules of $\pi_N$
in Table \ref{T:piNtr}. Then all $\ah_i$, $i\in I$, have the same
barb.}\end{Lem}
\begin{proof}It is obvious by the inference rules in the transitional semantics of $\pi_N$.
\end{proof}

In light of Lemma \ref{Lsa}, we can say that a transition group has
a barb, which is defined as the common barb arising from the actions
of probabilistic transitions in the transition group. A transition
group is called an {\it output transition group} (respectively, {\it
input transition group}) if its barb is of the form $\ol{x}$
(respectively, $x$).

\begin{table}\hspace{-0.8cm}
{\begin{tabular}{lll} & Out $\dfrac{\
}{\ol{x}y.P\{\xra{\ol{x}y_i}_{p_i}P\}_{i\in I}}$\quad $I=\{i:
p_i=p_x(y_i|y)>0\}$\quad
Tau  $\dfrac{\ }{\tau.P\{\os{\tau}\lra_{1}P\}}$\vspace{0.4cm} \\

& Inp $\dfrac{\ }{x(z).P\{\os{xy}\lra_{1}P\{y/z\}\}}$\hspace{4.5cm}
Mat
 $\dfrac{\pi.P\{\os{\ah_i}\lra_{p_i}P_i\}_{i\in
I}}{[x=x]\pi.P\{\os{\ah_i}\lra_{p_i}P_i\}_{i\in I}}$ \vspace{0.4cm} \\

& Sum-L $\dfrac{P\{\os{\ah_i}\lra_{p_i}P_i\}_{i\in
I}}{P+Q\{\os{\ah_i}\lra_{p_i}P_i\}_{i\in I}}$ \hspace{4.2cm} Par-L
 $\dfrac{P\{\os{\ah_i}\lra_{p_i}P_i\}_{i\in
I}}{P|Q\{\os{\ah_i}\lra_{p_i}P_i|Q\}_{i\in I}}$
\vspace{0.4cm} \\

& Comm-L $\dfrac{P\{\xra{\ol{x}y_i}_{p_i}P_i\}_{i\in
I'}\uplus\{\xra{\ol{x}(y_i)}_{p_i}P_i\}_{i\in I''}\quad
Q\{\os{xy_i}{\lra}_1Q_i\}}{P|Q\{\os{\tau}{\lra}_{p_i}P_i|Q_i\}_{i\in
I'} \uplus\{\os{\tau}{\lra}_{p_i}(\nu y_i)(P_i|Q_i)\}_{i\in I''}}$
 \vspace{0.4cm} \\

& Res-Out $\dfrac{P\{\xra{\ol{x}y_i}_{p_i}P_i\}_{i\in
I'}\uplus\{\xra{\ol{x}(y_i)}_{p_i}P_i\}_{i\in I''}}{(\nu
y_j)P\{\xra{\ol{x}y_i}_{p_i}\!(\nu y_j)P_i\}_{i\in
I'\backslash\{j\}}\uplus\{\xra{\ol{x}(y_i)}_{p_i}\!(\nu
y_j)P_i\}_{i\in
I''\backslash\{j\}}\uplus\{\xra{\ol{x}(y_j)}_{p_j}\!P_j\}}$ \;
$y_j\neq x$\vspace{0.4cm}\\

& Res-Inp $\dfrac{P\{\os{xy}\lra_{1}P'\}}{(\nu
z)P\{\os{xy}\lra_{1}(\nu z)P'\}}$  \quad $z\neq x,y$\hspace{2.2cm}
Res-Tau  $\dfrac{P\{\os{\tau}\lra_{p_i}P_i\}_{i\in I}}{(\nu
z)P\{\os{\tau}\lra_{p_i}(\nu z)P_i\}_{i\in I}}$ \vspace{0.4cm}\\

& Open-Inp $\dfrac{P\{\os{xy}\lra_{1}P'\}}{(\nu
y)P\{\os{xy}\lra_{1}P'\}}$  \quad $y\neq x$\hspace{2.95cm} Rep-Act
 $\dfrac{P\{\os{\ah_i}\lra_{p_i}P_i\}_{i\in I}}{!P\{\os{\ah_i}\lra_{p_i}P_i|!P\}_{i\in I}}$\vspace{0.4cm}\\

& Rep-Comm $\dfrac{P\{\xra{\ol{x}y_i}_{p_i}P_i\}_{i\in
I'}\uplus\{\xra{\ol{x}(y_i)}_{p_i}P_i\}_{i\in I''}\quad
P\{\os{xy_i}{\lra}_1P'_i\}}{!P\{\os{\tau}{\lra}_{p_i}P_i|P'_i|!P\}_{i\in
I'} \uplus\{\os{\tau}{\lra}_{p_i}(\nu y_i)(P_i|P'_i)|!P\}_{i\in
I''}}$\vspace{0.2cm}
\end{tabular}}
\caption{Transition group rules of $\pi_N$\label{T:piNtgr}}
\end{table}

Using Lemma \ref{Lsa}, the transition rules in Table \ref{T:piNtr}
are grouped in Table \ref{T:piNtgr}.

For later need, we introduce one more notation. A function $\mu$
from $\Omega$ to the closed unit interval $[0,1]$ is called a {\it
probability distribution} on $\Omega$ if $\sum_{x\in\Omega}\mu(x) =
1$. By $\mathcal{D}(\Omega)$ we denote the set of all probability
distributions on the set $\Omega$. If
$\mu\in\mathcal{D}(\Omega\times\Gamma),\ \omega\in\Omega$, and
$S\subseteq\Gamma$, we define $\mu(\omega,S)=\sum_{s\in
S}\mu(\omega,s)$.

Observe that each transition group, say
$P\{{\os{\ah_i}\lra}_{p_i}Q_i\}_{i\in I}$, gives rise to a
probabilistic distribution $\mu$ on $Act\times Proc$, where $\mu$ is
defined by
\begin{displaymath} \mu(\ah,Q)=\left\{
\begin{array}{ll}
p, & \textrm{if $P{\os{\ah}\lra}_{p}Q$ belongs to }P\{{\os{\ah_i}\lra}_{p_i}Q_i\}_{i\in I}\\
0, & \textrm{otherwise}.
\end{array} \right.
\end{displaymath}
Therefore, we sometimes write $P\lra\mu$ or
$P\xra{\mbox{barb}(\ah_i)}\mu$ for
$P\{{\os{\ah_i}\lra}_{p_i}Q_i\}_{i\in I}$, where
$\mbox{barb}(\ah_i)$ is the barb of the transition group.

We end this subsection with two properties of transition groups. The
first one corresponds to the image-finiteness of transition
relations in $\pi$. Here, by image-finiteness we mean that for any
process $P$ and action $\ah$ in $\pi$, there are only finitely many
processes $Q$ such that $P{\os{\ah}\lra}Q$.
\begin{Lem}\label{LImFin}{\rm Keep Convention \ref{Conv}. Then for every $\ah\in Act$ and $P\in Proc$, the set
$$\{\mu|_{\{\ah\}\times Proc}: P\lra\mu\}$$ is finite, where  $\mu|_{\{\ah\}\times Proc}$ is the
restriction of $\mu$ to $\{\ah\}\times Proc$.
}\end{Lem}
\begin{proof}It follows immediately from Lemmas \ref{Lfo} and \ref{Lbo}.
\end{proof}

The other property of transition groups is concerned with applying
substitution to transitions. To state it, we need the next
definition.
\begin{Def}{\rm\

$(1)$ A substitution $\sm$ is said to be {\it consistent} with
weakly bound names of a process $P$ if $z\sm\in\mbox{wbn}(P\sm)$
implies $z\not\in\mbox{fn}^*(P)$.

$(2)$ A substitution $\sm$ is said to be {\it compatible} with a
channel $x\in{\bf N}$ if for any $y,z\in{\bf N}$ it holds that
$$p_{x\sm}(u|y\sm)=\sum_{z\sm=u}p_{x}(z|y).$$
 }\end{Def}

We remark that the definition of compatibility here is a special
case of Definition 1 in \cite{Ying05}. A transition group under a
consistent and compatible substitution has the following property.

\begin{Prop}{\rm Let $P\{\os{\ah_i}\lra_{p_i}P_i\}_{i\in I}$ be a transition group.
Suppose that $\sm$ is a substitution satisfying the following
conditions:
\begin{itemize}
    \item[1)] $\sm$ is consistent with weakly bound names of all processes appearing
    in the inference for deriving the transition group;
    \item[2)] $\sm$ is compatible with the subjects of output actions appearing
    in the inference for deriving the transition group.
\end{itemize}
Then, there is a transition group
$P\sm\{\os{\ah_i\sm}\lra_{p_i}P_i\sm\}_{i\in I}$.
}\end{Prop}
\begin{proof}
It follows from a case by case check of all possible transition
groups $P\{\os{\ah_i}\lra_{p_i}P_i\}_{i\in I}$. The condition $1)$
is used to compute probability of $P\sm\os{\ah_i\sm}\lra P_i\sm$,
and the condition $2)$ is required when restricting $P$. We only
consider the case of Out and omit the remainder. In the case of Out,
we may assume that $P=\ol{x}y.R$, $p_x(y_i|y)=p_i$,
$\ah_i=\ol{x}y_i$, and $P_i=R$. Then the transition group is
$\ol{x}y.R\{\os{\ol{x}y_i}\lra_{p_i}R\}_{i\in I}$. This gives a
transition group
$\ol{x\sm}y\sm.R\sm\{\os{\ol{x\sm}u_j}\lra_{q_j}R\sm\}_{j\in J}$ by
the definition of substitution, where $J=\{j:
q_j=p_{x\sm}(u_j|y\sm)>0\}$. It follows from the condition $2)$ that
$$
q_j=p_{x\sm}(u_j|y\sm)=\sum_{y_i\sm=u_j}p_x(y_i|y)=\sum_{i\in
I_j}p_x(y_i|y)=\sum_{i\in I_j}p_i,
$$ where $I_j=\{i\in I: y_i\sm=u_j\}$. Therefore,
\begin{eqnarray*}
P\sm\{\os{\ah_i\sm}\lra_{p_i}P_i\sm\}_{i\in
I}&=&\ol{x\sm}y\sm.R\sm\{\os{\ol{x\sm}y_i\sm}\lra_{p_i}R\sm\}_{i\in
I}\\
&=&\ol{x\sm}y\sm.R\sm\{\os{\ol{x\sm}u_j}\lra_{\sum_{i\in I_j}p_i}R\sm\}_{j\in J}\\
&=&\ol{x\sm}y\sm.R\sm\{\os{\ol{x\sm}u_j}\lra_{q_j}R\sm\}_{j\in J},
\end{eqnarray*}as desired.
\end{proof}

\section{Barbed equivalence}
$\indent$Having built the transition group rules, we can now turn to
several behavioral equivalences in $\pi_N$. As mentioned in the last
section, collecting the probabilistic transitions arising from a
noisy channel into a group can alleviate some confusions, so the
transition group provides a useful way of defining behavioral
equivalences. Of course, other ways which do not use the transition
group are possible (for example, Definition 4 in \cite{Ying05}).
Recall that in $\pi_N$ two kinds of actions, noisy free output
$\ol{x}y$ and noisy bound output $\ol{x}(y)$, have the same barb
$x$. Clearly, not all observers can detect whether an emitted name
is bound. It is therefore natural to ask what is the effect of
reducing this discriminatory power. In this section, the concepts
about behavioral equivalences are based upon the assumption that the
observer is limited to seeing whether an action is enabled on a
given channel; behavioral equivalences with a more powerful observer
will be studied in the next section. Because the behavior of a
process in $\pi_N$ is evidently dependent on the noise probability
distribution, it is better to parameterize a behavioral equivalence
with the noise. However, for simplicity we assume that all processes
in a definition of behavioral equivalence are considered under the
same noisy environment, that is, the channel matrix of every name is
fixed.

Let us start with some basic notions. The {\it transitive closure}
of a binary relation $\mc{R}$ on $Proc$ is the minimal transitive
relation $\mc{R}^*$ on $Proc$ that contains $\mc{R}$, that is, if
$(P,Q)\in\mc{R}^*$, then there exist $P_0,\ldots,P_n\in Proc$
satisfying that $P=P_0$, $Q=P_n$, and $(P_{i-1},P_{i})\in\mc{R}$ for
$i=1,\ldots,n$. Thus, if $\mc{R}$ and $\mc{S}$ are two equivalence
relations on $Proc$, then so is $(\mc{R}\cup\mc{S})^*$. It turns out
that $(\mc{R}\cup\mc{S})^*$ is the smallest equivalence relation
containing both $\mc{R}$ and $\mc{S}$. For any equivalence relation
$\mc{R}$ on $Proc$, we denote by $Proc/\mc{R}$ the set of
equivalence classes induced by $\mc{R}$.

The definition below only takes the internal action into account.
\begin{Def}{\rm
An equivalence relation $\mathcal{R}$ on $Proc$ is a {\it reduction
bisimulation} if whenever $(P,Q)\in\mathcal{R}$, $P\os{\tau}\lra\mu$
implies $Q\os{\tau}\lra\eta$ for some $\eta$ satisfying
$\mu(\tau,C)=\eta(\tau,C)$ for any $C\in Proc/\mc{R}$. }\end{Def}

Note that the above reduction bisimulation and also other subsequent
notions on behavioral equivalences are defined in the same style as
Larsen-Skou's probabilistic bisimulation \cite{LarS91}, that is, the
probabilities of reaching an equivalence class have to be computed.
Recall that in a non-probabilistic setting we simply require that
$P\os{\tau}\lra P'$ implies $Q\os{\tau}\lra Q'$ and
$(P',Q')\in\mathcal{R}$, that is, it is enough that the agent $Q$
has a possibility to imitate the step of the agent $P$. However, if
we work in a probabilistic setting, in addition to need that the
second agent is able to imitate the first one, it is also reasonable
to require that he does it with the same probability. In other
words, we have to consider all the possible ways to imitate the
execution of the action, and thus we have to add the probabilities
associated with these possibilities. Correspondingly, summing up the
probabilities of reaching an equivalence class replaces the
condition $(P',Q')\in\mathcal{R}$ in a non-probabilistic setting. It
should be noted that the treatment of probabilities here is slightly
different from that of $\lambda$-bisimulation in \cite{Ying05},
where a higher probability of an action is allowed to simulate the
same action. In the literature, lumping equivalence, a notion on
Markov chains to aggregate state spaces \cite{Buc94,Hil96}, was also
defined by summing up the probabilities of reaching an equivalence
class; it needs not to consider the labels of transitions and is
different from our definitions on behavioral equivalences in
$\pi_N$.

Because the union of equivalence relations may not be an equivalence
relation, unlike in $\pi$, the union of reduction bisimulations is
not a reduction bisimulation in general. Nevertheless, we have the
following result.
\begin{Prop}\label{PLargredbisim}{\rm
Let $\bumpeq=(\bigcup_i\mathcal{R}_i)^*$, where $\mathcal{R}_i$ is a
reduction bisimulation on $Proc$. Then $\bumpeq$ is the largest
reduction bisimulation on $Proc$. }\end{Prop}
\begin{proof}
It suffices to show that $\bumpeq$ is a reduction bisimulation on
$Proc$. Suppose that $(P,Q)\in\bumpeq$ and $P\os{\tau}\lra\eta_0$.
Then there are $P_0,\ldots,P_n\in Proc$ and reduction bisimulations
$\mc{R}_{1'},\ldots,\mc{R}_{n'}$ such that $P=P_0$, $Q=P_n$, and
$(P_{i-1},P_{i})\in\mc{R}_{i'}$ for $i=1,\ldots,n$. By definition,
there exist $\eta_1,\ldots,\eta_{n}$ such that
$P_i\os{\tau}\lra\eta_i$ and
$\eta_{i-1}(\tau,C_{i'})=\eta_i(\tau,C_{i'})$ for $i=1,\ldots,n$ and
any $C_{i'}\in Proc/\mc{R}_{i'}$. Note that for any $C\in
Proc/\bumpeq$ and $\mc{R}_{i'}$, it follows from
$\mc{R}_{i'}\subseteq\bumpeq$ that $C=\bigcup_jC_{i'j}$ for some
$C_{i'j}\in Proc/\mc{R}_{i'}$, and moreover, $\bigcup_jC_{i'j}$ is a
disjoint union since every $C_{i'j}$ is an equivalence class. We
thus have that
\begin{eqnarray*}
\eta_{i-1}(\tau,C)&=&\eta_{i-1}(\tau,\bigcup_jC_{i'j}) \\
&=&\sum_j\eta_{i-1}(\tau,C_{i'j})\\
&=&\sum_j\eta_{i}(\tau,C_{i'j})\\
&=&\eta_{i}(\tau,C)
\end{eqnarray*}
for each $i=1,\ldots,n$. This means that
$\eta_{0}(\tau,C)=\eta_{n}(\tau,C)$. Hence, $\bumpeq$ is a reduction
bisimulation, as desired.
\end{proof}

The reduction bisimulation  $\bumpeq$ is called {\it reduction
bisimilarity}. In other words, $P$ and $Q$ are {\it reduction
bisimilar} if $(P,Q)\in\mathcal{R}$ for some reduction bisimulation
$\mathcal{R}$.

As a process equivalence, reduction bisimilarity is seriously
defective. For example, it relates any two processes that have no
internal actions, such as $\ol{x}a$ and ${\ol{y}a}$. To obtain a
satisfactory process equivalence, it is therefore necessary to allow
more to be observed of processes. In the sequel, we use $\theta$ to
rang over barbs but $\tau$. Based upon Lemma \ref{Lsa}, we have the
following definition.

\begin{Def}{\rm
The {\it observability predicate} $\downarrow_\theta$ in the
$\pi_N$-calculus is defined as follows:

$(1)$ $P\downarrow_a$ if $P$ has an input transition group with
subject $a$.

$(2)$ $P\downarrow_{\ol{a}}$ if $P$ has an output transition group
with subject $a$. }\end{Def}

Taking observability into consideration, we modify the notion of
reduction bisimulation as follows.

\begin{Def}\label{DBarBis}{\rm
An equivalence relation $\mathcal{R}$ on $Proc$ is a {\it (strong)
barbed bisimulation} if whenever $(P,Q)\in\mathcal{R}$,

$(1)$ $P\downarrow_\theta$ implies $Q\downarrow_\theta$;

$(2)$ $P\os{\tau}\lra\mu$ implies $Q\os{\tau}\lra\eta$ for some
$\eta$ satisfying $\mu(\tau,C)=\eta(\tau,C)$ for any $C\in
Proc/\mc{R}$.}\end{Def}

Analogous to Proposition \ref{PLargredbisim}, we have the following
fact.
\begin{Prop}\label{PLargbarbisim}{\rm
Let $\dot{\sim}=(\bigcup_i\mathcal{R}_i)^*$, where $\mathcal{R}_i$
is a barbed bisimulation on $Proc$. Then $\dot{\sim}$ is the largest
barbed bisimulation on $Proc$. }\end{Prop}
\begin{proof}
  Since barbed bisimulations are reduction bisimulations, we see
  that $\dot{\sim}$ is a reduction bisimulation by Proposition
  \ref{PLargredbisim}. For any $(P,Q)\in\dot{\sim}$, it is clear that
  $P\downarrow_\theta$ implies $Q\downarrow_\theta$. Thereby, $\dot{\sim}$ is
  a barbed bisimulation, and moreover, it is the largest
  one since it includes all barbed bisimulations on $Proc$.
\end{proof}

The barbed bisimulation $\dot{\sim}$ is called {\it (strong) barbed
bisimilarity}; we say that $P$ and $Q$ are {\it (strong) barbed
bisimilar}, denoted $P\dot{\sim}Q$, if $(P,Q)\in\mathcal{R}$ for
some barbed bisimulation $\mathcal{R}$. It follows readily from
definition that barbed bisimilarity is properly included in
reduction bisimilarity. In addition, the fact below is also obvious.

\begin{Lem}\label{Lcongbarb}{\rm
  If $P\equiv Q$, then $P\dot{\sim}Q$.
}\end{Lem}

Like reduction bisimilarity, barbed bisimilarity is not satisfactory
as a process equivalence as well. Nevertheless, it will underpin two
good relations, barbed equivalence and barbed congruence. Let us
begin with barbed equivalence.

\begin{Def}{\rm Two processes $P$ and $Q$ in $\pi_N$ are called {\it
(strong) barbed equivalent}, denoted $P\dot\approx Q$, if
$P|R\dot{\sim}Q|R$ for any $R$.}\end{Def}

It follows directly from the above definition that
$\dot\approx\subseteq\dot{\sim}$. On the other hand, there are a
large number of counter-examples to show that
$\dot{\sim}\nsubseteq\dot\approx$. For example, if $P\os{\rm
def}{=}\ol{x}a.\ol{y}b$ and $Q\os{\rm def}{=}\ol{x}a$, then for any
channel matrix of $x$, we have that $P\dot{\sim}Q$ by definition.
However, if one takes $R\os{\rm def}{=}x(w)$, then there exists $P|R
\{\os{\tau}{\lra}_{1}\ol{y}b\}$, and moreover, it cannot be matched
by $Q|R$ because the unique transition group $Q|R
\{\os{\tau}{\lra}_{1}\bf{0}\}$ having $\tau$ as barb yields $\ol{y}b
\dot\nsim \bf{0}$. Hence, $P|R\dot\approx Q|R$ does not hold.

Further, we have the following fact that will be of use later.

\begin{Lem}\label{LResCong}
  {\rm \

$(1)$ If $P\dot\approx Q$, then $(\nu z)(P|R)\dot\approx (\nu
z)(Q|R)$ for any $R$ and $z$.

$(2)$ Let $\mc{S}$ be an equivalence relation included in
$\dot{\sim}$. If for any $R$ and $z$, $(P,Q)\in\mc{S}$ implies
$((\nu z)(P|R), (\nu z)(Q|R))\in\mc{S}$, then
$\mc{S}\subseteq\dot\approx$.
  }
\end{Lem}
\begin{proof}For $(1)$, assume that $P\dot\approx Q$. By definition, we see that $P|R\dot\approx Q|R$ for any
$R$. Moreover, it is straightforward to show that $(\nu
z)P\dot\approx (\nu z)Q$ for any $z$. Therefore, the assertion $(1)$
holds.

For $(2)$, suppose that $(P,Q)\in\mc{S}$. Then we see by the
hypothesis of $\mc{S}$ that for any $R$,
$(P|R,Q|R)\in\mc{S}\subseteq\dot\sim$. Hence, $P|R\dot\sim Q|R$ for
any $R$. So $P\dot\approx Q$, and thus $\mc{S}\subseteq\dot\approx$,
finishing the proof.
\end{proof}

Finally, we introduce the concept of barbed congruence.
\begin{Def}\label{DBarbCong}{\rm Two processes $P$ and $Q$ in $\pi_N$ are {\it
(strong) barbed congruent}, denoted $P\dot\simeq Q$, if
$\mc{C}[P]\dot{\sim}\mc{C}[Q]$ for every process context
$\mc{C}$.}\end{Def}

In other words, two terms are barbed congruent if the agents
obtained by placing them into an arbitrary context are barbed
bisimilar. The following remark clarifies the relationship between
barbed equivalence and barbed congruence
\begin{Rem}\label{RBarnEquCong}
{\rm By definition, we see that barbed congruent processes in the
$\pi_N$-calculus are barbed equivalent, namely,
$\dot\simeq\subseteq\dot\approx$. In fact, this inclusion is strict.
The following example serves:

Let us take $P\os{\rm def}{=}x(w).[w=y]\tau$ and $Q\os{\rm
def}{=}x(w).[w=z]\tau$, and suppose that all communication channels,
except for $x$, are noiseless. We also assume that the channel
matrix of $x$ is given by
\begin{equation*}
\begin{split}
&p_x(y|y)=0.5,\quad p_x(z|y)=0.5;\\
&p_x(y|z)=0.5,\quad p_x(z|z)=0.5;\\
&p_x(s|s)=1 \mbox{ for any }s\neq y,z.
\end{split}
\end{equation*}

It follows readily that for any $R\in Proc$ with
$R\{\os{\ah_i}{\lra}_{p_i}R'_i\}_{i\in I}$, if
$\ah_i\not\in\{\ol{x}y,\ol{x}z,\ol{x}(y),\ol{x}(z)\}$, then
$P|R\dot\sim Q|R$. Moreover, if there exists
$\ah_i\in\{\ol{x}y,\ol{x}z,\ol{x}(y),\ol{x}(z)\}$, then any
transition group of $R$ having $\ah_i$ as an action must be one of
the following forms:
\begin{align*}
&(1)\quad R\{\os{\ol{x}y}{\lra}_{0.5}R'_1\}\uplus\{\xra{\ol{x}z}_{0.5}R'_1\};\\
&(2)\quad R\{\os{\ol{x}(y)}{\lra}_{0.5}R'_2\}\uplus\{\xra{\ol{x}z}_{0.5}(\nu y)R'_2\};\\
&(3)\quad R\{\os{\ol{x}y}{\lra}_{0.5}(\nu z)R'_3\}\uplus\{\xra{\ol{x}(z)}_{0.5}R'_3\};\\
&(4)\quad R\{\os{\ol{x}(y)}{\lra}_{0.5}(\nu
z)R'_4\}\uplus\{\xra{\ol{x}(z)}_{0.5}(\nu y)R'_4\}.
\end{align*}
For the form $(1)$, we have that
\begin{align*}
&P|R\{\os{\tau}{\lra}_{0.5}\tau|R'_1\}\uplus\{\os{\tau}{\lra}_{0.5}R'_1\} \mbox{ and }\\
&Q|R\{\os{\tau}{\lra}_{0.5}\tau|R'_1\}\uplus\{\os{\tau}{\lra}_{0.5}R'_1\},
\end{align*}which means that $P|R\dot\sim Q|R$.
For the form $(2)$, we have that
\begin{align*}
&P|R\{\os{\tau}{\lra}_{0.5}(\nu y)(\tau|R'_2)\}\uplus\{\os{\tau}{\lra}_{0.5}(\nu y)R'_2\} \mbox{ and }\\
&Q|R\{\os{\tau}{\lra}_{0.5}\tau|(\nu
y)R'_2\}\uplus\{\os{\tau}{\lra}_{0.5}(\nu y)R'_2\}.
\end{align*}
This yields that $P|R\dot\sim Q|R$ because of $(\nu
y)(\tau|R'_2)\equiv\tau|(\nu y)R'_2$. The form $(3)$ is similar to
that of the form $(2)$, and we can also get that $P|R\dot\sim Q|R$.
For the form $(4)$, we see that
\begin{align*}
&P|R\{\os{\tau}{\lra}_{0.5}(\nu y)(\tau|(\nu z)R'_4)\}\uplus\{\os{\tau}{\lra}_{0.5}(\nu y, z)R'_4\} \mbox{ and }\\
&Q|R\{\os{\tau}{\lra}_{0.5}(\nu z)(\tau|(\nu
y)R'_4)\}\uplus\{\os{\tau}{\lra}_{0.5}(\nu y, z)R'_4\}.
\end{align*}Because $(\nu y)(\tau|(\nu z)R'_4)
\equiv\tau|(\nu y, z)R'_4\equiv(\nu z)(\tau|(\nu y)R'_4)$, we get
that $P|R\dot\sim Q|R$. Summarily, we have that $P|R\dot\sim Q|R$
for any $R$, and thus $P\dot\approx Q$.

On the other hand, let $\mc{C}=x(y).[\ ]|\ol{x}s.\ol{x}s$. Then
$\mc{C}[P]=x(y).x(w).[w=y]\tau|\ol{x}s.\ol{x}s$ and
$\mc{C}[Q]=x(y).x(w).[w=z]\tau|\ol{x}s.\ol{x}s$. It is easy to check
that $\mc{C}[P]\dot\nsim \mc{C}[Q]$, and thus $P\dot\simeq Q$ does
not hold, as desired.
 \hfill$\square$ }\end{Rem}

\section{Bisimilarity}
$\indent$As mentioned in the last section, behavioral equivalences
under a powerful observer that can differentiate between noisy free
output and noisy bound output are investigated in this section.

We begin with a classical notion, bisimulation.
\begin{Def}\label{DBis}{\rm
An equivalence relation $\mathcal{R}$ on $Proc$ is a {\it (strong)
bisimulation} if whenever $(P,Q)\in\mathcal{R}$, $P\lra\mu$ implies
$Q\lra\eta$ for some $\eta$ satisfying $\mu(\ah,C)=\eta(\ah,C)$ for
any $\ah\in Act$ and $C\in Proc/\mc{R}$.}\end{Def}

The next result gives the largest bisimulation.
\begin{Prop}\label{PLargbisim}{\rm
Let $\sim=(\bigcup_i\mathcal{R}_i)^*$, where $\mathcal{R}_i$ is a
bisimulation on $Proc$. Then $\sim$ is the largest bisimulation on
$Proc$. }\end{Prop}
\begin{proof}
  It is similar to that of Proposition \ref{PLargredbisim}.
\end{proof}

The largest bisimulation $\sim$ is called {\it (strong)
bisimilarity}. In other words, $P$ and $Q$ are {\it (strong)
bisimilar}, written $P\sim Q$, if $(P,Q)\in\mathcal{R}$ for some
bisimulation $\mathcal{R}$. As an immediate consequence of
Definitions \ref{DBarBis} and \ref{DBis}, we have the following.

\begin{Lem}\label{LBisBar}
  {\rm Any two bisimilar processes are barbed bisimilar, i.e., $\sim \subseteq\dot{\sim}$.
  }
\end{Lem}

The remark below tells us that bisimilarity is not included in
barbed congruence.
\begin{Rem}
{\rm Just like in the $\pi$-calculus, bisimilar processes in the
$\pi_N$-calculus may not be barbed congruent, that is,
$\sim\nsubseteq\dot\simeq$. The following counter-example serves:
Let $P\os{\rm def}{=}\ol{a}u|b(v)$ and $Q\os{\rm
def}{=}\ol{a}u.b(v)+b(v).\ol{a}u$, and suppose, for simplicity, that
there is a channel $x$ with $p_x(b|b)=1$. Then we see that $P\sim
Q$. However, the context $\mc{C}\os{\rm def}{=}(x(a).[\ ])|\ol{x}b$
yields that $\mc{C}[P]\dot\nsim \mc{C}[Q]$. Therefore, $P\dot\simeq
Q$ does not hold.
 \hfill$\square$ }\end{Rem}

The following notion will provide an equivalent characterization of
bisimilarity.
\begin{Def}{\rm A family of binary relations $\sim_n$ on $Proc$ stratifying
the bisimilarity is defined inductively as follows:
\begin{enumerate}
    \item[(1)] $\sim_0$ is the universal relation on processes.
    \item[(2)] For any $0<n<\infty$, $(P,Q)\in\sim_n$ if

    (2.1) $P\lra\mu$ implies
$Q\lra\eta$ for some $\eta$ satisfying
$\mu(\ah,C_{n-1})=\eta(\ah,C_{n-1})$ for any $\ah\in Act$ and
$C_{n-1}\in Proc/\sim_{n-1}$, and

(2.2) $Q\lra\eta$ implies $P\lra\mu$ for some $\mu$ satisfying
$\mu(\ah,C_{n-1})=\eta(\ah,C_{n-1})$ for any $\ah\in Act$ and
$C_{n-1}\in Proc/\sim_{n-1}$.
    \item[(3)] $(P, Q)\in\sim_\infty$ if $(P,Q)\in\sim_n$ for all $n<\infty$.
\end{enumerate}
}\end{Def}

Observe that every $\sim_n$ is an equivalence relation, so the
notation $Proc/\sim_{n-1}$ in the above definition makes sense. Note
also that $\sim_0, \sim_1,\ldots,\sim_\infty$ is a decreasing
sequence of relations.

The result below gives another way to check bisimilarity.
\begin{Prop}\label{PStrBisim}
 {\rm $P\sim Q$ if and only if $P\sim_\infty Q$.
}\end{Prop}
\begin{proof}We first prove the necessity. By definition, we only need to show that $\sim
\subseteq\sim_n$ for all $n<\infty$. Proceed by induction on $n$.
The case $n=0$ is trivial. Assume that $\sim \subseteq\sim_{n-1}$.
For any $(P,Q)\in\sim$ and $P\lra\mu$, it follows from the
definition of $\sim$ that there exists $\eta$ such that $Q\lra\eta$
and $\mu(\ah,C)=\eta(\ah,C)$ for any $\ah\in Act$ and $C\in
Proc/\sim$. By induction hypothesis $\sim \subseteq\sim_{n-1}$, we
see that $\mu(\ah,C_{n-1})=\eta(\ah,C_{n-1})$ for any $\ah\in Act$
and $C_{n-1}\in Proc/\sim_{n-1}$. Consequently, $\sim
\subseteq\sim_{n}$, as desired.

Now, let us show the sufficiency. It is enough to prove that
$\sim_\infty$ is a bisimulation. Suppose that $P\sim_\infty Q$ and
$P\lra\mu$. Then for each $n<\infty$, there is $\eta_n$ such that
$Q\lra\eta_n$ and $\mu(\ah,C_n)=\eta_n(\ah,C_n)$ for any $\ah\in
Act$ and $C_n\in Proc/\sim_n$. By Lemma \ref{LImFin}, the number of
distinct $\eta_n$'s is finite, so there is $\eta'$ such that
$\eta'=\eta_n$ for infinitely many $n$. Since $\sim_0,
\sim_1,\ldots,\sim_\infty$ is a decreasing sequence of equivalence
relations, we get that $\mu(\ah,C_n)=\eta'(\ah,C_n)$ for any $C_n\in
Proc/\sim_n$. This gives rise to $(P,Q)\in\sim_n$ for all
$n<\infty$, which means that $(P,Q)\in\sim_\infty$. Therefore,
$\sim_\infty$ is a bisimulation, finishing the proof.
\end{proof}

For later need, let us pause to develop a ``bisimulation up to''
technique. Firstly, we make the following definition.
\begin{Def}{\rm A binary symmetric relation $\mathcal{R}$ on $Proc$ is a {\it
bisimulation up to $\sim$} if whenever $(P,Q)\in\mathcal{R}$,
$P\lra\mu$ implies $Q\lra\eta$ for some $\eta$ satisfying
$\mu(\ah,C)=\eta(\ah,C)$ for any $\ah\in Act$ and $C\in
Proc/(\mc{R}\cup\sim)^*$. }\end{Def}

The following fact shows that any bisimulation up to $\sim$ relation
is a bisimulation, as expected.
\begin{Prop}\label{LBisUp}
  {\rm
If $\mathcal{R}$ is a bisimulation up to $\sim$, then
$\mathcal{R}\subseteq\sim$.
  }
\end{Prop}
\begin{proof}
Let $\mc{S}=(\mc{R}\cup\sim)^*$. For any $(P,Q)\in\mc{S}$, there
exist $P_0,\ldots,P_n$ such that $P_0=P$, $P_n=Q$, and
$(P_{i-1},P_i)\in\mc{R}\cup\sim$ for $i=1,\ldots,n$. If
$(P_{i-1},P_i)\in\mc{R}$, then $P_{i-1}\lra\mu_{i-1}$ implies
$P_i\lra\mu_i$ for some $\mu_i$ satisfying
$\mu_{i-1}(\ah,C)=\mu_i(\ah,C)$ for any $\ah\in Act$ and $C\in
Proc/\mc{S}$. If $(P_{i-1},P_i)\in\sim$, then $P_{i-1}\lra\mu_{i-1}$
implies $P_i\lra\mu_i$ for some $\mu_i$ satisfying
$\mu_{i-1}(\ah,C')=\mu_i(\ah,C')$ for any $\ah\in Act$ and $C'\in
Proc/\sim$. This, together with the fact $\sim\subseteq\mc{S}$,
yields that $\mu_{i-1}(\ah,C)=\mu_i(\ah,C)$ for any $\ah\in Act$ and
$C\in Proc/\mc{S}$. As a result, for any $P_0\lra\mu_0$, there are
$\mu_1,\ldots,\mu_n$ such that for $i=1,\ldots,n$, $P_i\lra\mu_i$
and $\mu_{i-1}(\ah,C)=\mu_i(\ah,C)$ for any $\ah\in Act$ and $C\in
Proc/\mc{S}$. Therefore, $\mc{S}$ is a bisimulation, and thus
$\mathcal{R}\subseteq\sim$. This completes the proof.
\end{proof}

We now establish a bisimulation up to $\sim$ relation which will be
used in the next section.
\begin{Lem}\label{LResBis}
  {\rm Let
 \begin{equation*}
\begin{split}
\mc{R}=\{((\nu \widetilde{z})(P|R),(\nu \widetilde{z})(Q|R)):
\widetilde{z}\mbox{ are arbitrary names}, \mbox{and }P,Q,R\in Proc
\mbox{ with } P\sim Q
 \}.
\end{split}
\end{equation*}
Then $\mc{R}$ is a bisimulation up to $\sim$. }
\end{Lem}
\begin{proof}Clearly, $\mc{R}$ is an equivalence relation. To check that it is a bisimulation up to $\sim$,
we appeal to a case analysis on the last rules applied in the
inference of related transitions. It needs to examine all inductive
steps through combinations of the transition group rules of Par,
Comm, Res-Out, Res-Inp, Res-Tau, and Open-Inp. This is a long and
routine argument, so we omit the details here.
\end{proof}

To obtain a congruence based on actions, we make the following
definition.
\begin{Def}{\rm Two processes $P$ and $Q$ in $\pi_N$ are called {\it
(strong) full bisimilar}, denoted $P\simeq Q$, if $P\sigma\sim
Q\sigma$ for any substitution $\sigma$.}\end{Def}

Like in $\pi$, bisimilarity in $\pi_N$ is not preserved by
substitution, as illustrated below.
\begin{Rem}\label{RBisFull}
{\rm It follows immediately from definition that full bisimilar
processes are necessarily bisimilar, i.e., $\simeq\subseteq\sim$.
Nevertheless, $\sim\nsubseteq\simeq$. In other words, there are
 some bisimilar processes that are not full bisimilar. For
example, let $P\os{\rm def}{=}\ol{a}u|b(v)$ and $Q\os{\rm
def}{=}\ol{a}u.b(v)+b(v).\ol{a}u$. Then for any noisy channels, we
always have that $P\sim Q$. However, the substitution $\sigma$
defined by
\begin{displaymath}
\sm(x)=\left\{ \begin{array}{ll}
a, & \textrm{if $x=b$}\\
x, & \textrm{otherwise}\qquad\qquad
\end{array} \right.
\end{displaymath}
gives rise to that $P\sm=\ol{a}u|a(v)$ and
$Q\sm=\ol{a}u.a(v)+a(v).\ol{a}u$. Obviously, $P\sm\nsim Q\sm$, and
thus $P$ and $Q$ are not full bisimilar. \hfill$\square$  }
\end{Rem}

\section{A hierarchy of behavioral equivalences}
$\indent$In the previous two sections, we have introduced several
behavioral equivalences. Some simple inclusion relationships among
them have been established. In this section, we consummate the
relationships and then give a hierarchy of these behavioral
equivalences.

The following theorem shows that bisimilar processes are barbed
equivalent.
\begin{Thm}\label{TBisBarEqu}{\rm For any two processes $P$ and $Q$ in $\pi_N$, if $P\sim Q$, then $P\dot\approx Q$.
}\end{Thm}
\begin{proof}By Lemmas \ref{LBisUp} and \ref{LResBis}, we see that $P\sim Q$ implies $(\nu z)(P|R)\sim (\nu z)(Q|R)$ for any
$R$ and $z$. Using the fact $\sim\subseteq\dot\sim$ obtained in
Lemma \ref{LBisBar}, we get from Lemma \ref{LResCong} $(2)$ that
$\sim\subseteq\dot\approx$, thus finishing the proof.
\end{proof}

\begin{Rem}\label{Rnoncoin}
{\rm In the $\pi$-calculus, it is well known that strong barbed
equivalence coincides with strong bisimilarity; see, for example,
Theorem 2.2.9 in \cite{SanW01a}. However, in the $\pi_N$-calculus
the converse of the above theorem is not true in general. In other
words, two barbed equivalent processes may not be bisimilar. For
example, let us consider the barbed equivalent processes $P\os{\rm
def}{=}x(w).[w=y]\tau$ and $Q\os{\rm def}{=}x(w).[w=z]\tau$ in
Remark \ref{RBarnEquCong}, and keep the hypothesis of related
channel matrices. Then we find that
$$P\{\os{xy}{\lra}_{1}\tau\} \mbox{ and } Q\{\os{xy}{\lra}_{1}\bf{0}\}.$$ This shows us that
$P\nsim Q$. Obviously, this non-coincidence of barbed equivalence
and bisimilarity arises from the noise of channel $x$.
\hfill$\square$ }
\end{Rem}

We continue to discuss the relationship between barbed congruence
and full bisimilarity. To this end, we need one more concept.
\begin{Def}{\rm
  An equivalence relation $\mc{R}$ on processes is said to be a {\it
  process congruence} if $(P,Q)\in\mc{R}$ implies
  $(\mc{C}[P],\mc{C}[Q])\in\mc{R}$ for every process context $\mc{C}$.}
\end{Def}

The following is an easy consequence, which is useful for checking
process congruence.
\begin{Prop}\label{PEleCont}{\rm
  An equivalence relation $\mc{R}$ is a process congruence if and
  only if it is preserved by all elementary contexts.}
\end{Prop}

The next observation gives a basic process congruence; its proof
follows immediately from the definition of structural congruence.
\begin{Prop}\label{PStrCong}{\rm
  The structural congruence $\equiv$ is a process congruence.}
\end{Prop}

As an immediate consequence of Definition \ref{DBarbCong},
Proposition \ref{PStrCong}, and Lemma \ref{Lcongbarb}, we have the
following.
\begin{Coro}{\rm
  If $P\equiv Q$, then $P\dot\simeq Q$.
}\end{Coro}

For subsequent need, we show that $\simeq$ is also a process
congruence.
\begin{Lem}\label{LFullCon}{\rm
 $\simeq$ is a process congruence, and moreover, it is the largest process congruence included in $\sim$.
}\end{Lem}
\begin{proof}
We first show that $\simeq$ is a process congruence. Suppose that
$P\simeq Q$, i.e., $P\sm\sim Q\sm$ for every substitution $\sm$. By
Proposition \ref{PEleCont}, we only need to prove that for every
elementary context $\mc{C}$ and any substitution $\sm$,
$\mc{C}[P\sm]\sim \mc{C}[Q\sm]$. For $\mc{C}=\pi.[\ ]+M$, if the
prefix $\pi$ is of form $\ol{x}y$, $\tau$, or $[x=y]\pi$, then it
follows directly from $P\sm\sim Q\sm$ that $\mc{C}[P\sm]\sim
\mc{C}[Q\sm]$. In the case $\mc{C}=x(z).[\ ]+M$, we see that
$\mc{C}[P\sm]=x(z).P\sm+M$ and $\mc{C}[Q\sm]=x(z).Q\sm+M$. Because
$z$ is not strongly bound in $P\sm$ or $Q\sm$, we get by $P\simeq Q$
that for any $z'$, $P\sm\{z'/z\}\sim Q\sm\{z'/z\}$. This gives rise
to $x(z).P\sm\sim x(z).Q\sm$, and thus $\mc{C}[P\sm]\sim
\mc{C}[Q\sm]$. By carrying out an analysis of the transition group
rules related to composition, restriction, and replication, it is
routine to check that $\mc{C}[P\sm]\sim \mc{C}[Q\sm]$ holds for the
other four elementary contexts, and we do not go into the details.

By definition, we see that $\simeq\subseteq\sim$. We now verify that
$\simeq$ is the largest process congruence included in $\sim$. Let
$\sim^*$ be an arbitrary process congruence included in $\sim$.
Suppose that $P\sim^* Q$ and $\sm=\{y_1,\ldots, y_n/x_1,\ldots,
x_n\}$. Without loss of generality, we assume that there is a
noiseless channel $s$ with $s\not\in\mbox{fn}^*(P\sm, Q\sm)$, and
set $$\mc{C}\os{\rm def}{=}(\nu s)(\ol{s}y_1.\ldots
.\ol{s}y_n|s(x_1).\ldots .s(x_n).[\ ]).$$ Then $\mc{C}[P]\sim^*
\mc{C}[Q]$, hence $\mc{C}[P]\sim \mc{C}[Q]$. Notice that
$\mc{C}[P]\{(\os{\tau}{\lra}_{1})^n(\nu s)P\sm\equiv P\sm\}$, where
$(\os{\tau}{\lra}_{1})^n$ is the $n$-fold composition of
$\os{\tau}{\lra}_{1}$, therefore
$\mc{C}[Q]\{(\os{\tau}{\lra}_{1})^n\sim P\sm\}$. But
$\mc{C}[Q]\{(\os{\tau}{\lra}_{1})^n(\nu s)Q\sm\equiv Q\sm\}$ only,
so $P\sm\sim Q\sm$. We thus get that $P\simeq Q$, and hence
$\sim^*\subseteq\simeq$, as desired.
\end{proof}

Based on the previous lemmas, we can prove the next result.
\begin{Thm}\label{TFullBisBarCon}{\rm For any two processes $P$ and $Q$ in $\pi_N$, if $P\simeq Q$, then $P\dot\simeq Q$.
}\end{Thm}
\begin{proof}We see from  Lemma \ref{LFullCon} that $\simeq$ is a process congruence included in $\sim$. Since
$\sim\subseteq\dot\approx$ by Theorem \ref{TBisBarEqu}, $\simeq$ is
a process congruence included in $\dot\approx$. By definition,
$\dot\simeq$ is the largest process congruence included in
$\dot\approx$, therefore we have that $\simeq\subseteq\dot\simeq$,
finishing the proof of the theorem.
\end{proof}

It would be expected that $\dot\simeq\subseteq\simeq$. However, this
is not true, as we shall see.
\begin{Rem}
{\rm Like Theorem \ref{TBisBarEqu}, the converse of the above
theorem is not true in general, that is, two barbed congruent
processes may not be full bisimilar. Even two barbed congruent
processes may not be bisimilar in the $\pi_N$-calculus. For example,
take $P\os{\rm def}{=}(\nu y)\ol{x}y.\ol{x}y$ and $Q\os{\rm
def}{=}(\nu y)\ol{x}y.(\nu y)\ol{x}y$, and suppose, for simplicity,
that the channel $x$ is noiseless. It is easy to check by induction
on context $\mc{C}$ that $\mc{C}[P]\dot\sim \mc{C}[Q]$ holds for any
context. Consequently, $P\dot\simeq Q$ by definition. Nevertheless,
notice that there is a transition group
$P\{\os{\ol{x}(y)}{\lra}_1\ol{x}y\}$ and the only transition group
of $Q$ making $P\sim Q$ possible is $Q\{\os{\ol{x}(y)}{\lra}_1(\nu
y)\ol{x}y\}$. But it is obvious that $\ol{x}y\nsim (\nu y)\ol{x}y$.
Hence, $P\nsim Q$. \hfill$\square$ }\end{Rem}

Finally, based on our results in Sections 4--6, we summarize the
hierarchy of the behavioral equivalences in the $\pi_N$-calculus,
which is depicted in Figure \ref{FHier}(b).
\begin{Thm}{\rm  In the $\pi_N$-calculus,

$(1)$
$\simeq\subseteq\dot\simeq\subseteq\dot\approx\subseteq\dot\sim\subseteq\bumpeq$
and $\simeq\subseteq\sim\subseteq\dot\approx$; each of the
inclusions can be strict.

$(2)$ Neither $\sim\subseteq\dot\simeq$ nor
$\dot\simeq\subseteq\sim$ holds.
  }
\end{Thm}

\section{Conclusion}
$\indent$This paper is devoted to a hierarchy of behavioral
equivalences in the $\pi_N$-calculus, the $\pi$-calculus with noisy
channels. First, we have developed an early transitional semantics
of the $\pi_N$-calculus and provided two presentations of the
transition rules. It is worth noting that this semantics is not a
directly translated version of the late semantics of $\pi_N$ in
\cite{Ying05}, and we have found that not all bound names are
compatible with alpha-conversion in the noisy environment, which is
a striking dissimilarity between $\pi_N$ and $\pi$. As a result, an
Open-Inp rule for inputting bound names is required. Then we have
introduced some notions of behavioral equivalences in $\pi_N$,
including reduction bisimilarity, barbed bisimilarity, barbed
equivalence, barbed congruence, bisimilarity, and full bisimilarity.
Some basic properties of them have also been stated. Finally, we
have established an integrated hierarchy of these behavioral
equivalences. In particular, because of the noisy nature of
channels, the coincidence of bisimilarity and barbed equivalence, as
well as the coincidence of full bisimilarity and barbed congruence,
in the $\pi$-calculus does not hold in $\pi_N$.

There are some limits and problems arising from the present work
which are worth further studying. We only present a hierarchy of
strong behavioral equivalences; the corresponding weak version that
ignores invisible internal actions is a research topic. Some
algebraic laws and axiomatizations of these behavioral equivalences
are interesting problems for future research. A hierarchy of
behavioral equivalences in some subcalculus (for example,
asynchronous $\pi$-calculus where asynchronous observers are less
discriminating than synchronous observers
\cite{AmaCS98,Bou92,FouG98,FouG05,HonT91}) is yet to be addressed.
Note that the converse statements of Theorems \ref{TBisBarEqu} and
\ref{TFullBisBarCon} cannot hold. Hence, a necessary and sufficient
condition for the converse statements to be true is desirable.
Finally, the reliability of processes in the $\pi_N$-calculus
initiated in \cite{Ying05} remains an interesting issue when using
the early transitional semantics of $\pi_N$ and non-approximate
bisimilarity.

\section*{Acknowledgment}
The author would like to thank Professor Mingsheng Ying for some
helpful discussions and invaluable suggestions.


\end{document}